\documentclass[final,5pt,onecolumn]{elsarticle}
\usepackage[margin=1.8cm]{geometry}% by courtesy of Mico
\usepackage[english]{babel}
\usepackage{amsmath,amssymb,graphicx,hyperref}
\usepackage{xcolor}
\usepackage[utf8]{inputenc}
\usepackage{listings}
\usepackage{xcolor}
\usepackage{hyperref}
\usepackage{array, longtable, tabularx}% added long table
\usepackage{pdflscape}% added
\hypersetup{colorlinks=true, citecolor=blue, filecolor=blue, linkcolor=blue, urlcolor=blue}
\definecolor{pastelyellow}{HTML}{FFFFDF}
\journal{Computer Physics Communications}

\begin{document}
\begin{frontmatter}

\title{VASPKIT: A User-friendly Interface Facilitating High-throughput Computing and Analysis Using VASP Code}

\author{Vei Wang}
\cortext[author]{\textit{E-mail address:} wangvei@icloud.com}
\address{Department of Applied Physics, Xi'an University of Technology, Xi'an 710054, China}

\author{Nan Xu}
\address{College of Chemical and Biological Engineering, Zhejiang University, Hangzhou 310027, China}  

\author{Jin-Cheng Liu}
\address{Department of Chemistry and Key Laboratory of Organic Optoelectronics \& Molecular Engineering of Ministry of Education, Tsinghua University, Beijing 100084, China} 

\author{Gang Tang}
\address{Theoretical Materials Physics, Q-MAT, CESAM, Universit$\acute{e}$ de Li$\grave{e}$ge, Li$\grave{e}$ge, Belgium} 

\author{Wen-Tong Geng}
\address{School of Materials Science \& Engineering, University of Science and Technology Beijing, Beijing 100083, China}

\begin{abstract}
We present the VASPKIT, a \textcolor{black}{command-line} program that aims at providing a powerful and user-friendly interface to perform high-throughput analysis of \textcolor{black}{a variety of} material properties from the raw data \textcolor{black}{produced by} the VASP code. It \textcolor{black}{consists of} mainly the pre- and post-processing modules. The former module is designed to prepare and manipulate input files such as the necessary input files generation, symmetry analysis, supercell transformation,  $k$-path generation for a given crystal structure. The latter module is designed to extract and analyze the raw data about elastic mechanics, electronic structure, charge density, electrostatic potential, linear optical coefficients, wave function plots in real space, and etc. This program can run conveniently in either interactive user interface or command line mode. The command-line options allow \textcolor{black}{the user} to perform high-throughput calculations together with bash scripts. This article gives an overview of the program structure and presents illustrative examples for some of its usages. \textcolor{black}{The program can run on Linux, MacOS, and Windows platforms. The executable versions of VASPKIT and  the related examples, together with the tutorials, are available in its official website \href{https://vaspkit.com}{vaspkit.com}}.

\end{abstract}

\end{frontmatter}

%\linenumbers
\noindent
{\bf PROGRAM SUMMARY}\\
\begin{small}
\noindent
{\em Program Title:} VASPKIT \\
{\em Licensing provisions:} GPLv3\\
{\em Programming language:} Fortran, Python\\
{\em Computer:} All computers with a Fortran compiler supporting at least Fortran 90.\\
{\em Operating system:} All operating systems with such a Fortran compiler.\\
{\em Nature of problem:} This program has the purpose of providing a powerful and user-friendly interface to perform high-throughout calculations together with the widely-used VASP code.\\
{\em Solution method:} VASPKIT is able to extract, calculate and even plot the mechanical, electronic, optical and magnetic properties from density functional calculations together with bash and python scripts. It can run \textcolor{black}{in} either interactive user interface or command line mode.\\
\end{small}

\section{Introduction}
\textcolor{black}{With the rapid development of high-performance computations and computational algorithms,  high-throughput computational analysis and discovery of materials has become an emerging research field because it promises to avoid time-consuming try and error experiments and explore the hidden potential behind thousands of potentially unknown materials within short timeframes that the real experiments might take a long time.}
Density functional theory (DFT) is one of the most popular methods that can treat both model systems and realistic materials in a quantum mechanical way \cite{Hohenberg1964,Kohn1965,Payne1992,Jones1989,Jones2015}. It is not only used to understand the observed behavior of solids, including the structural, mechanical, electronic, magnetic and optical properties, but increasingly more to predict characteristics of compounds that have not yet been determined experimentally \cite{Sato2010,Dietl2014,Jain2016,Freysoldt2014,Pokluda2015,Zhang2017a,Oganov2019}. 

The last two decades have witnessed tremendous progress in the \textcolor{black}{methodology} development for first-principles calculations of materials properties. There are dozens of electronic-structure \textcolor{black}{computation} packages \textcolor{black}{that} have been developed based on DFT so far, such as Abinit \cite{Gonze2016}, CASTEP \cite{Clark2005}, VASP \cite{Kresse1996,Kresse1996a}, Siesta \cite{Soler2002}, Quantum Espresso \cite{Giannozzi2009,Giannozzi2017}, Elk \cite{Dewhurst2016} and WIEN2k \cite{Blaha2020}, \textcolor{black}{with great success in}  exploring material properties. One of the common features for these packages is that post-processing is required to plot or extract into a human-readable format from the raw data. \textcolor{black}{There are two popular commercial programs, Materials Studio \cite{ms2020} and QuantumATK \cite{Smidstrup2019}, providing graphical user interface (GUI) that allows the researchers to easily build, visualize, and review results and calculation setup up with a set of mouse actions. However, these GUI programs become less productive when the users want to perform batch processing operations. In contrast, several open-source post-processing packages, such as Python Materials Genomics (pymatgen) \cite{Ong2013}, Atomic Simulation Environment (ASE) \cite{Larsen2017}, and PyProcar \cite{Herath2020} provide powerful command-line interfaces to efficiently extract, plot and analyze the raw data in batch mode but require the users to be proficient in Python programming language. 
It is worth mentioning here that both $lev00$ \cite{Kantorovich2020} and qvasp \cite{Yi2020} are two interactive menu-driven programs written in Fortran which mainly focus on the post-processing of electronic structure calculations using VASP and other codes.}

In this article we will introduce a toolkit, here referred to as VASPKIT which is developed for providing a powerful and user-friendly integrated input/output environment tool to perform initial setup for calculations and post-processing analysis to derive various material properties from the raw data calculated using the VASP code. It is capable of calculating the elastic, electronic, optical and catalytic properties including equation of state, elastic constants, carrier effective masses, fermi surfaces, band structure unfolding for supercell models, linear optical coefficients, joint density of states, transition dipole moment, wave functions plots in real space, thermo energy correction,  and etc. In addition, it also allows \textcolor{black}{the users} to perform high-throughput calculations with low barriers to entry. For example, we recently performed high-throughput calculations to screen hundreds of two-dimensional (2D) semiconductors from near 1000 monolayers using this program together with VASP \cite{Wang2018}. The VASPKIT \textcolor{black}{remains in development, with growing functionality,  and is ready to be extended to work directly with outputs from other electronic structure packages}.

The rest of this paper is organized as follows: In Section 2  the workflow and basic features of the pre-processing module as implemented into VASPKIT are described. Section 3 presents the computational algorithms and some examples illustrating the capabilities of post-processing module in the VASPKIT code. Finally,  a short summary is given in Summary.

\section{\textcolor{black}{Capabilities of the}  Pre-Processing Module}
\begin{figure}[htbp]
\centering
\includegraphics[scale=0.4]{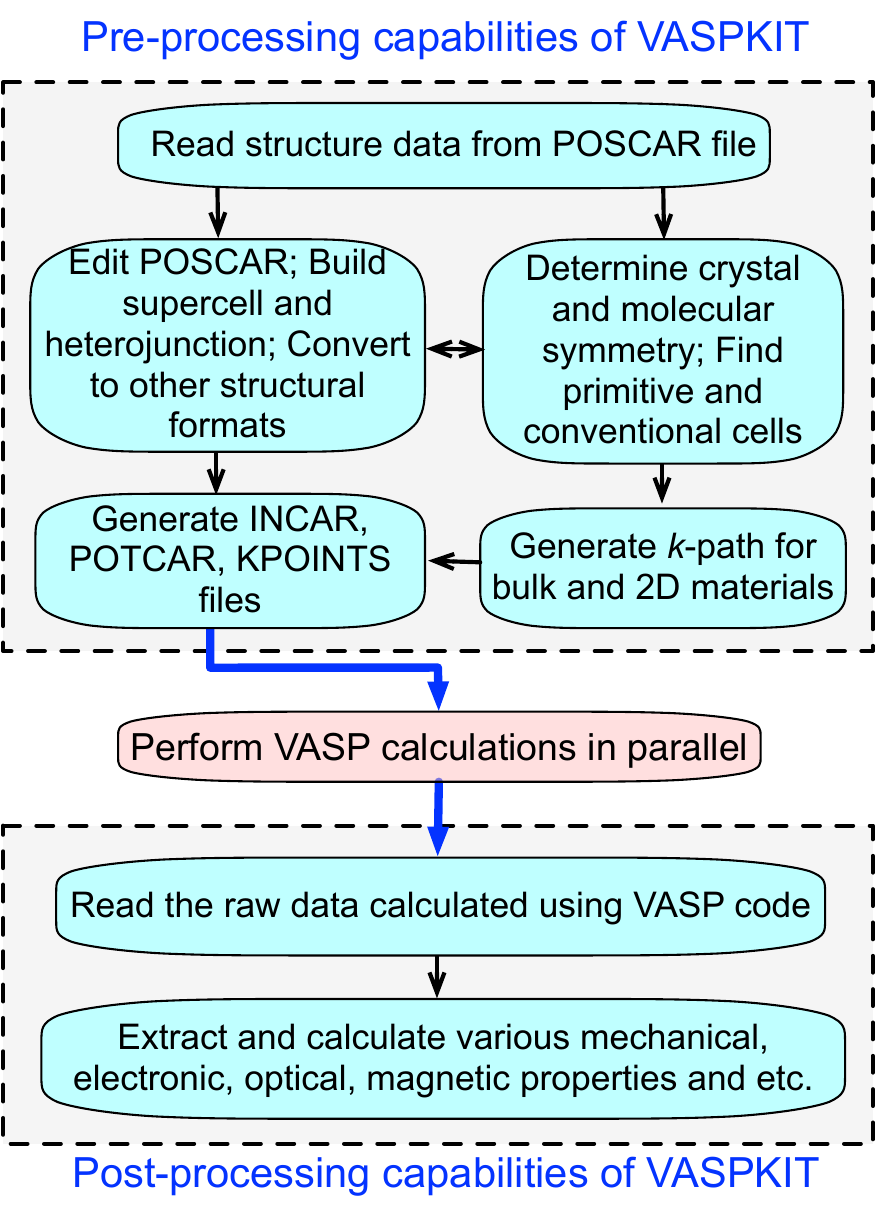}
\caption{\label{pre-proc}(Color online) A structural overview of VASPKIT package.}
\end{figure}

The workflow of VASPKIT package is illustrated in Fig. \ref{pre-proc}. In the pre-processing module, the program first reads the POSCAR file and then prepares the rest three input files (INCAR, POTCAR and KPOINTS) to perform  DFT calculations using VASP. It can also manipulate the structure file such as building supercell, generating the suggested $k$-path for band structure calculation,
determining the crystal symmetry information, or finding the conventional/primitive cell for a given lattice by employing the symmetry analysis library Spglib \cite{Togo2018}. Furthermore, it can convert POSCAR to several widely-used structural formats, such as XCrysDen  (.xsf) \cite{Kokalj2003}, Crystallographic Information Framework (.cif) \cite{Hall1991} or Protein Data Bank (.pdb) formats \cite{Bernstein1977}.

\subsection{\textcolor{black}{Definitions and conversions of crystal structures}} 
The crystal structures are often provided by basis vectors and point coordinates of labeled atoms. Lattice basis vectors $\mathbf{A}$ are represented by three row vectors

\begin{equation}
\mathbf{A}=\left(\begin{array}{l}{\mathbf{a}} \\ {\mathbf{b}} \\ {\mathbf{c}}\end{array}\right)=\left(\begin{array}{lll}{a_{x}} & {a_{y}} & {a_{z}} \\ {b_{x}} & {b_{y}} & {b_{z}} \\ {c_{x}} & {c_{y}} & {c_{z}}\end{array}\right).
\end{equation}

The position of an ion is represented by  a row vector either in fractional coordinates ($x$, $y$, $z$) with respect to basis vector lengths or in Cartesian coordinates ($X$, $Y$, $Z$). \textcolor{black}{The relationship of these two  coordinates is written as}

\begin{equation}
\left(\begin{array}{l}{X} \\ {Y} \\ {Z}\end{array}\right)=\mathbf{A}^T\left(\begin{array}{l}{x} \\ {y} \\ {z}\end{array}\right)=\left(\begin{array}{lll}{a_{x}} & {b_{x}} & {c_{x}} \\ {a_{y}} & {b_{y}} & {c_{y}} \\ {a_{z}} & {b_{z}} & {c_{z}}\end{array}\right)\left(\begin{array}{l}{x} \\ {y} \\ {z}\end{array}\right),
\end{equation}
\textcolor{black}{where $\mathbf{A}^T$ denotes the matrix transpose of lattice basis vectors $\mathbf{A}$.}

\begin{figure}[htbp]
\centering
\includegraphics[scale=0.5]{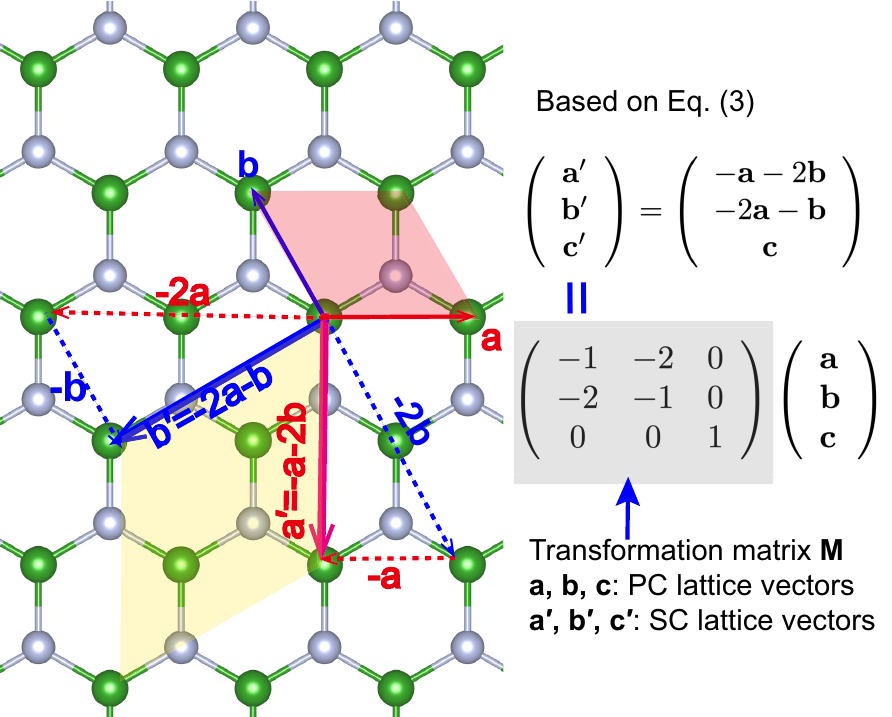}
\caption{\label{cell}(Color online) \textcolor{black}{Schematic illustration of how to build a supercell from the  lattice vectors of primitive cell (PC) and the specified transformation matrix. The supercell and primitive cell are indicated by the yellow and red rhombuses.}}
\end{figure}

The \textcolor{black}{conversion from one lattice basis ($\mathbf{a}$, $\mathbf{b}$, $\mathbf{c}$)} to another choice of lattice basis ($\mathbf{a}'$, $\mathbf{b}'$, $\mathbf{c}'$) is given by
\begin{equation}\label{AMa}
\left(\begin{array}{l}{\mathbf{a}'} \\ {\mathbf{b}'} \\ {\mathbf{c}'}\end{array}\right)=\mathbf{M} \cdot\left(\begin{array}{l}{\mathbf{a}} \\ {\mathbf{b}} \\ {\mathbf{c}}\end{array}\right),
\end{equation}
\textcolor{black}{where $\mathbf{M}$ is the transformation matrix. Its determinant $|\mathbf{M}|$ defines the ratio between the supercell and primitive cell volumes in the real space.  Figure \ref{cell} shows how to construct a supercell (SC) from the specified transformation matrix and the primitive cell (PC) lattice vectors.}

\subsection{\textcolor{black}{Generation of suggested $k$-path}} 

\begin{figure}[htbp]
\centering
\includegraphics[scale=0.5]{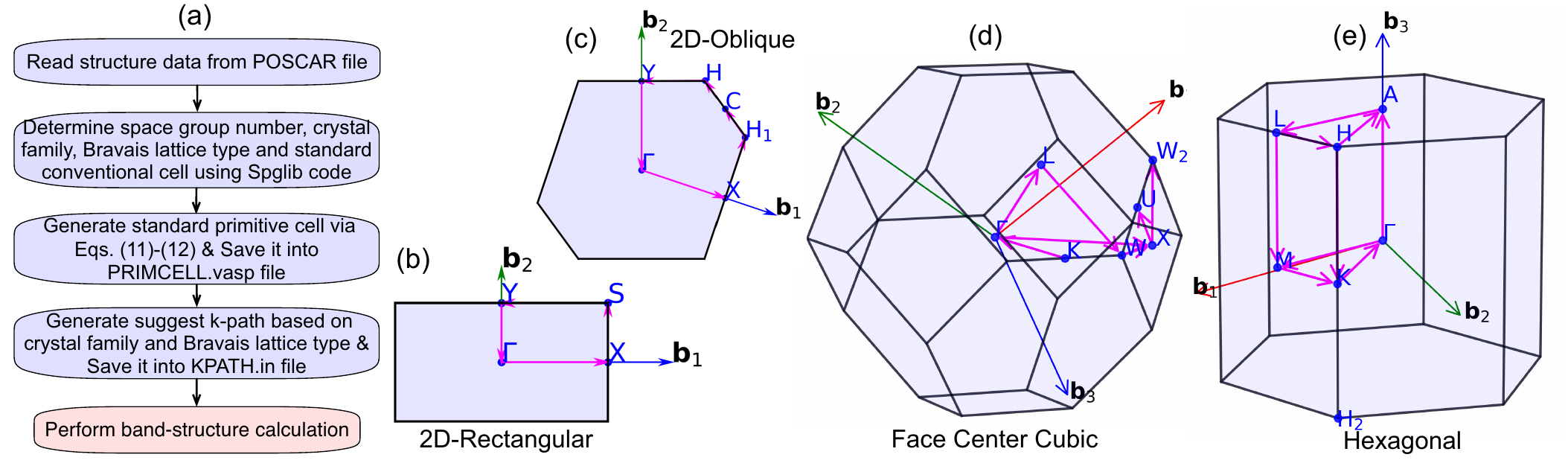}
\caption{\label{vbz}(Color online) (a) Workflow of the algorithm used in the $k$-path utility. The first Brillouin zone, special high symmetry points, and recommended $k$-paths for (a) 2D rectangular, (b) 2D oblique,  (c)  face-centered cubic and (d)  hexagonal close packed lattices respectively.}
\end{figure}

In order to plot a band structure, one needs to define a set of $k$-points \textcolor{black}{along desired high-symmetry directions in the Brillouin zone (BZ)}. The $k$-path utility automatically generates the suggested $k$-path for a given 2D \cite{Wang2018} or bulk \cite{Hinuma2017} crystal structure.  \textcolor{black}{The flowchart of the algorithm to determine the suggested $k$-path for a given crystal is shown in Figure \ref{vbz} (a). Specifically, VASPKIT first determines the space group number, crystal family and Bravais lattice type from the input structure, typically read from the POSCAR file; a standard conventional cell is then identified and constructed by idealizing the lattice vectors based on the axial lengths and the interaxial angles, aiming to eliminate the non-unique choices in the possible shapes of BZ in certain Bravais lattices \cite{Setyawan2010,Hinuma2017};  then the standard primitive cell is determined by transforming the basis vectors of the standard conventional cell according to Eq. (\ref{c2m}),}

\begin{equation}\label{c2m}
\left(\begin{array}{l}\mathbf{a}_p \\ \mathbf{b}_p \\ \mathbf{c}_p\end{array}\right)=\mathbf{P} \cdot\left(\begin{array}{l}\mathbf{a}_c \\ \mathbf{b}_c \\ \mathbf{c}_c\end{array}\right),
\end{equation}
 \textcolor{black}{where ($\mathbf{a}_p$, $\mathbf{b}_p$, $\mathbf{c}_p$) and  ($\mathbf{a}_c$, $\mathbf{b}_c$, $\mathbf{c}_c$) are the basis vectors of primitive and conventional systems, respectively,  $\mathbf{P}$  is the transformation matrix from the standardized conventional cell to the primitive cell, as discussed in Table 3 in Ref. \cite{Hinuma2017}, and the subscripts $c$ and $p$ represent the primitive and conventional cells respectively. The atomic position of an ion in fractional coordinates transformed from the basis vectors of conventional cell to those of primitive cell is written as below: }
\begin{equation}
\left(\begin{array}{l}x_{p} \\y_{p} \\z_{p}
\end{array}\right)=\mathbf{P}^{-1}\left(\begin{array}{l}x_{c} \\y_{c} \\z_{c}
\end{array}\right).
\end{equation}

\textcolor{black}{It should be noted that the number of atoms in the PC is generally less than that in SC. This implies that the transformation from SC to PC leads to some duplicated atoms which must be removed. In the final step, the $k$-path utility automatically saves the standard primitive cell and the suggested  $k$-path into the PRIMCELL.vasp and KPATH.in files respectively. In addition to the automatic generation of the suggested $k$-path when a crystal structure is given as input}, VASPKIT also provides the python script to visualize the specified $k$-path in the first Brillouin zone using Matplotlib plotting library \cite{Hunter2007}.  As illustrative examples, the recommended $k$-paths of 2D-rectangular, 2D-oblique and face-centered cubic and hexagonal lattices are show in Figs. \ref{vbz} (b)-(e) respectively.

\section{\textcolor{black}{Capabilities of the} Post-Processing Module} 
\begin{figure}[htbp]
\centering
\includegraphics[scale=0.36]{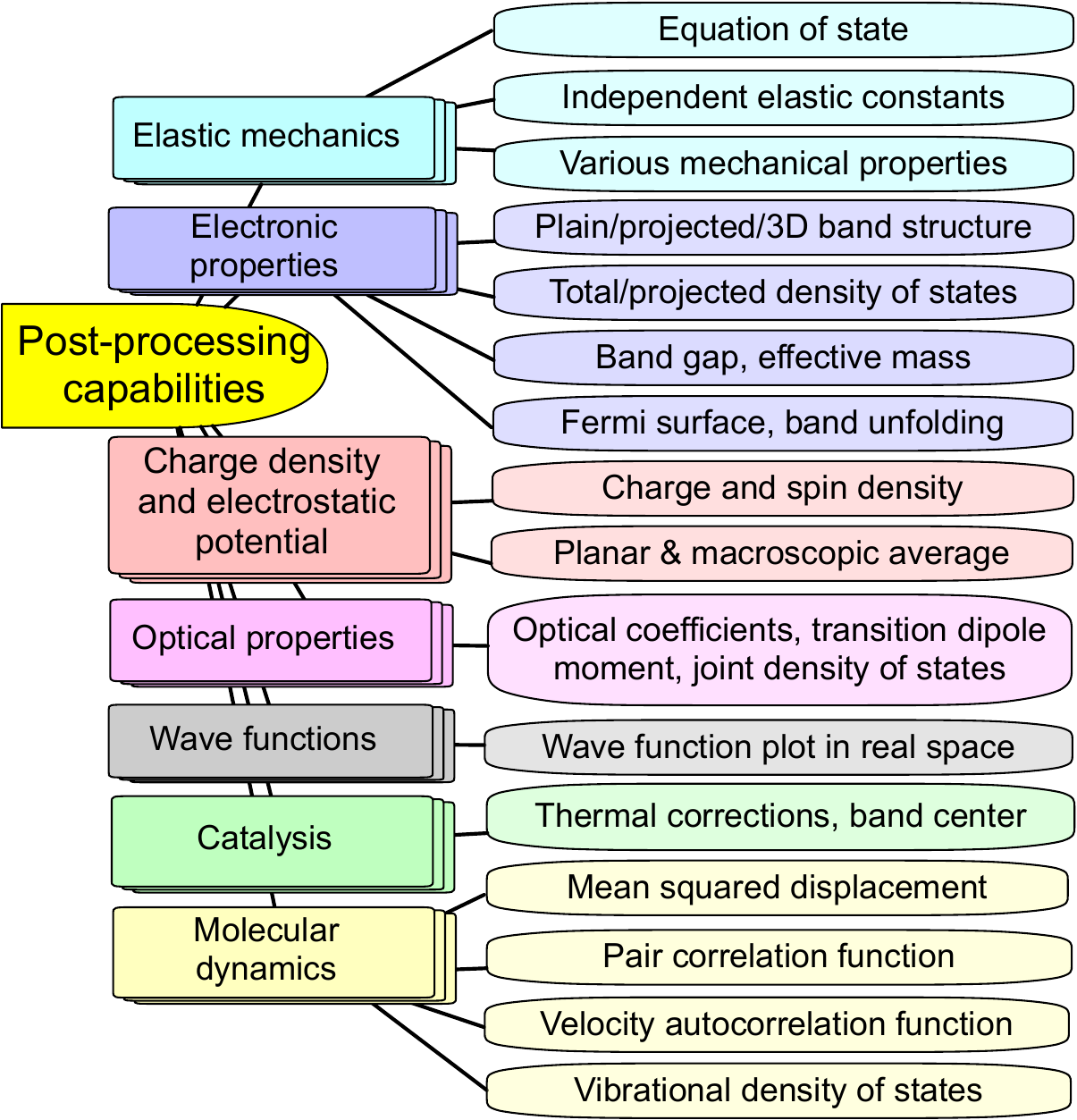}
\caption{\label{Post-pro}(Color online) A structural overview of the post-processing module implemented into the VASPKIT package.}
\end{figure}

Figure \ref{Post-pro} displays an overview of the post-processing features as implemented into the VASPKIT package. This module is designed to extract and analyze the raw data including elastic mechanics, electronic, charge density, electrostatic potential, optical wave-function, catalysis and molecular dynamics related properties. We next present the computational algorithms and some examples to illustrate the capabilities of the post-processing module.

\subsection{Elastic mechanics}
The second-order elastic constants (SOECs) play a crucial role in \textcolor{black}{governing} the mechanical and dynamical properties of materials, especially on the stability and stiffness. Within the linear elastic region, the stress $\mathbf{\sigma} = \left(\sigma_{1}, \sigma_{2}, \sigma_{3}, \sigma_{4}, \sigma_{5}, \sigma_{6}\right)$ response of solids to external loading strain $\boldsymbol{\varepsilon} = \left(\varepsilon_{1}, \varepsilon_{2}, \varepsilon_{3}, \varepsilon_{4}, \varepsilon_{5}, \varepsilon_{6}\right)$ satisfies the generalized Hooke's law and can be simplified in the Voigt notation \cite{Voigt1928},

\begin{equation}\label{stress-strain}
\sigma_{i}=\sum_{j=1}^{6} \text{C}_{i j} \varepsilon_{j},
\end{equation}
where strain $\sigma_{i}$ and stress $\varepsilon_{j}$ are represented as a vector with 6 independent components respectively, i.e., $1 \leq i,j \leq 6$. C$_{ij}$ is the second order elastic stiffness tensor expressed by a 6 $\times$ 6 symmetric matrix in units of GPa. The elastic stiffness tensor C$_{ij}$ can be determined \textcolor{black}{using} the first-order derivative of the stress-strain curves proposed by Nielsen and Martin \cite{Nielsen1983,Nielsen1985}, as expressed in Eq. (\ref{stress-strain}). The number of independent elastic constants depends on the symmetry of \textcolor{black}{the} crystal. The lower the symmetry means the more the independent elastic constants. For example, the cubic crystals have three but the triclinic ones have 21 independent elastic constants. The classification of the different crystal system with the corresponding number of independent elastic constants for bulk materials are summarized in Table \ref{iec_3d} \cite{Zhang2017,Golesorkhtabar2013,Nye1985}.  

\begin{table*}[hbt]
\fontsize{9}{11}\selectfont
\setlength{\tabcolsep}{1.5pc}
\caption{Classification of crystal systems, point group classes and space-group number are provided with the number of independent second elastic constants for bulk materials. In the last column, several prototype materials are shown.}
\label{iec_3d}
\begin{tabular*}{\textwidth}{@{}l@{\extracolsep{\fill}}ccccc}
\hline
Crystal system & Point groups & Space-groups & Number of  independent SOECs  & Material prototypes \\\hline
Triclinic      & 1, $\overline{1}$ &1-2 & 21  & - \\
Monoclinic     & $m, 2, \frac{2}{m}$ & 3-15 & 13 & ZrO$_2$ \\
Orthorhombic   & $222, m m 2, \frac{2}{m} \frac{2}{m} \frac{2}{m}$ & 16-74 & 9  & TiS$_2$ \\
Tetragonal I   & $422,4 m m, \overline{4}2 m, \frac{4}{m} \frac{2}{m} \frac{2}{m}$ & 89-142 & 6 & MgF$_2$ \\  
Tetragonal II  & $4, \overline{4}, \frac{4}{m}$ & 75-88 & 7 & CaMoO$_4$\\
Trigonal I     & $32,3 m, \overline{3} \frac{2}{m}$ & 149-167 & 6   & $\alpha$-Al$_2$O$_3$\\
Trigonal II    & $3, \overline{3}$ & 143-148 & 7   & CaMg(CO$_3$)$_2$\\
Hexagonal      & $622,6 m m, \overline{6} 2 m, \frac{6}{m} \frac{2}{m} \frac{2}{m}, 6, \overline{6}, \frac{6}{m}$  & 168-194  & 5 & Ti \\
Cubic          & $432, \overline{4} 3 m, \frac{4}{m} \overline{3} \frac{2}{m}, 23, \frac{2}{m} \overline{3}$  & 195-230  & 3 &Diamond  \\\hline
\end{tabular*}
\end{table*}

An alternative theoretical approach to calculate elastic constants is based on the energy variation by applying small strains to the equilibrium lattice configuration \cite{LePage2001}. The elastic energy $\Delta E \left(V,\left\{\varepsilon_{i}\right\}\right)$ of a solid under the harmonic approximation is given by

\begin{equation}\label{elastic_energy}
\begin{aligned}\Delta E\left(V,\left\{\varepsilon_{i}\right\}\right)=E\left(V,\left\{\varepsilon_{i}\right\}\right)-E\left(V_{0}, 0\right) \\ =\frac{V_{0}}{2} \sum_{i, j=1}^{6} \text{C}_{i j} \varepsilon_{j} \varepsilon_{i}\end{aligned},
\end{equation}
\textcolor{black}{where $E\left(V_{0}, 0\right)$ and $E\left(V,\left\{\varepsilon_{i}\right\}\right)$  are the total energies of the equilibrium and distorted lattice cells, with the volume of $V_{0}$ and $V$}, respectively. In the energy-strain method the elastic stiffness tensor is derived from the second-order derivative of the total energy versus strain curves \cite{LePage2001}. In general, the stress-strain method requires higher computational \textcolor{black}{precision} to achieve the same accuracy \textcolor{black}{than} the energy-strain method. Nevertheless, the former requires much smaller set of distortions than the latter \cite{Zhang2017,Golesorkhtabar2013,Yu2010,LePage2002,LePage2001}. Considering that the energy-strain \textcolor{black}{relation} has less stress sensitivity than the stress-strain one, the former method has been implemented into the VASPKIT package. Meanwhile, the determination of elastic stability criterion is also provided in the elastic utility based on the necessary and sufficient elastic stability conditions in the harmonic approximation \cite{Hashin1962} for various crystal systems proposed by Mouhat et al \cite{Zhang2017,Golesorkhtabar2013,Mouhat2014}.

\textcolor{black}{When a crystal is deformed by applying strain $\boldsymbol{\varepsilon}$, the relation} of lattice vectors between the distorted and equilibrium cells is given by  

\begin{equation}
\label{distortedcell}
\left(\begin{array}{l}{\mathbf{a}^{\prime}} \\ {\mathbf{b}^{\prime}} \\ {\mathbf{c}^{\prime}}\end{array}\right)=\left(\begin{array}{l}{\mathbf{a}} \\ {\mathbf{b}} \\ {\mathbf{c}}\end{array}\right) \cdot(\mathbf{I}+\boldsymbol{\epsilon}),
\end{equation}
where $\mathbf{I}$ is  the 3 $\times$ 3 identity matrix. The strain tensor $\boldsymbol{\epsilon}$ is defined by 

\begin{equation}
\label{strain_tensor}
\boldsymbol{\epsilon}=\left(\begin{array}{lll}
\varepsilon_{1} & \varepsilon_{6} / 2 & \varepsilon_{5} / 2 \\
\varepsilon_{6} / 2 & \varepsilon_{2} & \varepsilon_{4} / 2 \\
\varepsilon_{5} / 2 & \varepsilon_{4} / 2 & \varepsilon_{3}
\end{array}\right).
\end{equation}

\begin{figure}[htbp]
\centering
\includegraphics[scale=0.4]{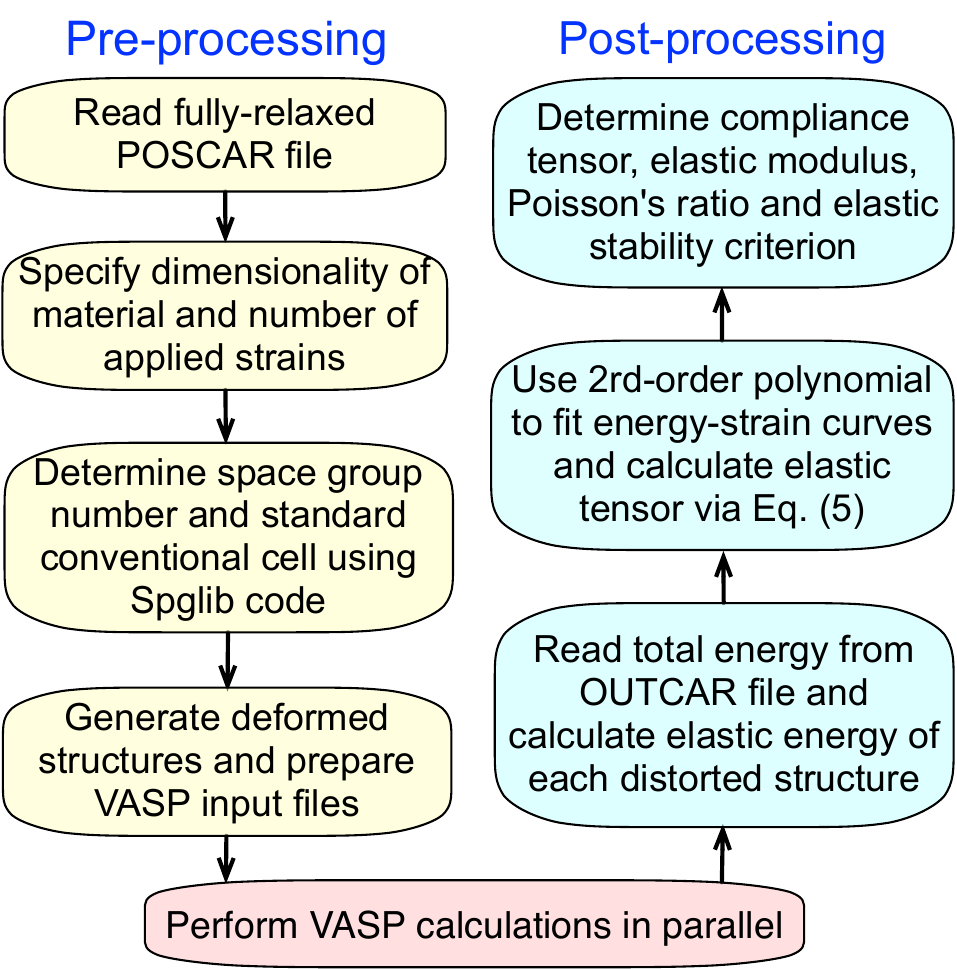}
\caption{\label{elastic}(Color online) Workflow of the algorithm to determine the second-order elastic constants based on energy-strain method used in the elastic utility.}
\end{figure}

The workflow of elastic utility is shown in Fig. \ref{elastic}. \textcolor{black}{VASPKIT first reads the equilibrium structure from POSCAR in which both lattice parameters and ﻿atomic positions are fully relaxed. In addition, the dimensionality of material (either 2D or 3D) and number of applied strain $\boldsymbol{\varepsilon}$ need to be specified as input. For 2D materials, in order to avoid mirror interactions the periodic slabs are required to separate by sufficiently large vacuum layer in $c$ direction. In the second step, the space group number and the type of input structure are analyzed by using the Spglib code \cite{Togo2018} to determine how many independent elastic constants need to be calculated. A classification of the different crystal system with the corresponding number of independent elastic constants is given in Table \ref{iec_3d}. Furthermore,  a standard conventional cell needs to be adopted in the following calculations since the the components of C$_{i j}$  are dependent on the choice of coordinate system and lattice vectors. After that, based on the determined space group number, a series of distorted structure with specified values of strain around the equilibrium are generated via Eq. (\ref{distortedcell}). Next, the elastic energies are calculated for each distorted structure by using VASP. Then, a polynomial fitting procedure is applied to calculate the second derivative at equilibrium of the energy with respect to the strain. Finally, various mechanical properties such as bulk, shear modulus and Poisson's ratio for polycrystalline materials are determined. }

\textcolor{black}{We take the cubic structure as an example to demonstrate how to calculate its independent elastic constants by using the energy-strain method. For cubic system,  the three independent elastic constants C$_{11}$, C$_{12}$ and C$_{44}$, are expressed in an elastic stiffness tensor matrix}

\begin{equation}
\label{cubice}
C_{ij}^{cubic}=\left(\begin{array}{cccccc}
C_{11} & C_{12} & C_{12} & 0 & 0 & 0 \\
C_{12} & C_{11} & C_{12} & 0 & 0 & 0 \\
C_{12} & C_{12} & C_{11} & 0 & 0 & 0 \\
0 & 0 & 0 & C_{44} & 0 & 0 \\
0 & 0 & 0 & 0 & C_{44} & 0 \\
0 & 0 & 0 & 0 & 0 & C_{44}
\end{array}\right).
\end{equation}

\textcolor{black}{After substituting  Eq. (\ref{cubice}) into Eq. (\ref{elastic_energy}), the elastic energy is written as below:}

\begin{equation}
\begin{aligned}
\frac{\Delta E}{V}=\frac{1}{2} &\left(C_{11} \varepsilon_{1} \varepsilon_{1}+C_{11} \varepsilon_{2} \varepsilon_{2}+C_{11} \varepsilon_{3} \varepsilon_{3}+C_{12} \varepsilon_{1} \varepsilon_{2}+C_{12} \varepsilon_{1} \varepsilon_{3}+C_{12} \varepsilon_{2} \varepsilon_{1}\right.\\
&+C_{12} \varepsilon_{2} \varepsilon_{3}+C_{12} \varepsilon_{3} \varepsilon_{1}+C_{12} \varepsilon_{3} \varepsilon_{2}+C_{44} \varepsilon_{4} \varepsilon_{4}+C_{44} \varepsilon_{5} \varepsilon_{5}+C_{44} \varepsilon_{6} \varepsilon_{6}).
\end{aligned}
\end{equation}

\textcolor{black}{When applied the tri-axial shear strain $\boldsymbol{\varepsilon}$=(0,0,0,$\delta$,$\delta$,$\delta$), Eq. (\ref{cubice}) becomes }

\begin{equation}\
\label{cubic1}
\frac{\Delta E}{V}=\frac{3}{2}C_{44} \delta^{2}.
\end{equation}

\textcolor{black}{Similarly, $C_{11}$+$C_{12}$  can be obtained by using the strain $\boldsymbol{\varepsilon}$=($\delta$,$\delta$,0,0,0,0):}

\begin{equation}
\label{cubic2}
\frac{\Delta E}{V}=\left(C_{11}+C_{12}\right) \delta^{2}.
\end{equation}

\textcolor{black}{Also, $C_{11}+2 C_{12}$ is calculated using the strain $\boldsymbol{\varepsilon}$=($\delta$,$\delta$,$\delta$,0,0,0):}

\begin{equation}
\label{cubic3}
\frac{\Delta E}{V}=\frac{3}{2}\left(C_{11}+2 C_{12}\right) \delta^{2}.
\end{equation}

\textcolor{black}{In order to calculate the elastic stiffness constants given above, the elastic energies of a set of deformed configurations in the distortion range -2\% $\leq$ $\delta$ $\leq$ +2\% with an increment of 0.5\% are investigated using VASP. After that, the quadratic coefficients are determined by fitting the energy versus distortion relationship, and finally the second order elastic constants $C_{ij}$ are determined by solving the equations (\ref{cubic1})-(\ref{cubic3}) during the post-processing of elastic utility. The details of strain modes and the derived elastic constants for each crystal system based on energy-strain approach are listed in Appendix A.}  

For polycrystalline materials, the crystallites are randomly \textcolor{black}{oriented}, and such materials can be considered to be quasi-isotropic or isotropic in a statistical sense. Thus, the bulk modulus $K$ and shear modulus $G$ are generally obtained by averaging the single-crystal elastic constants. Three of the most widely used averaging approaches have been implemented into the elastic utility: Voigt \cite{Voigt1928}, Reuss \cite{Reuss1929} and Hill \cite{Hill1952} schemes. Hill has shown that the Voigt and Reuss elastic moduli are the strict upper and lower \textcolor{black}{bounds}  \cite{Hill1952}, respectively. The arithmetic mean of the Voigt and Reuss bounds termed the Voigt-as Reuss-Hill (VRH) average is found \textcolor{black}{to be} better approximation to the actual elastic behavior of a polycrystalline material. 

\textcolor{black}{The} Voigt bounds are given by the following equations:
\begin{equation}
\left\{\begin{array}{l}
{9 K_\text{V}=\left(\text{C}_{11}+\text{C}_{22}+\text{C}_{33}\right)+2\left(\text{C}_{12}+\text{C}_{23}+\text{C}_{31}\right)} \\
{15 G_\text{V}=\left(\text{C}_{11}+\text{C}_{22}+\text{C}_{33}\right)-\left(\text{C}_{12}+\text{C}_{23}+\text{C}_{31}\right)} \\
{\quad+4\left(\text{C}_{44}+\text{C}_{55}+\text{C}_{66}\right)}
\end{array}\right.,
\end{equation}
while the Reuss bounds are given by:
\begin{equation}
\left\{\begin{array}{c}
{1 / K_\text{R}=\left(\text{S}_{11}+\text{S}_{22}+\text{S}_{33}\right)+2\left(\text{S}_{12}+\text{S}_{23}+\text{S}_{31}\right)} \\
{15 / G_\text{R}=4\left(\text{S}_{11}+\text{S}_{22}+\text{S}_{33}\right)-4\left(\text{S}_{12}+\text{S}_{23}+\text{S}_{31}\right)} \\
{\quad+3\left(\text{S}_{44}+\text{S}_{55}+\text{S}_{66}\right)}
\end{array}\right.,
\end{equation}
\textcolor{black}{where S$_{ij}$ are the components of compliance tensor, which correspond to the matrix elements of the inverse of the elastic tensor, namely, $\left[S_{i j}\right]=\left[C_{i j}\right]^{-1}$. The} Voigt and Reuss bounds are rigorous upper and lower bounds of $K$ and $G$, respectively. Based on \textcolor{black}{the} Voigt and Reuss bounds, Hill defined $K_\text{VRH}=1/2\left(K_\text{V}+K_\text{R}\right)$ and $G_\text{VRH}=1/2\left(G_\text{V}+G_\text{R}\right)$, known as the Voigt-Reuss-Hill average.\cite{Hill1952} \textcolor{black}{Using} the values of bulk modulus $K$ and shear modulus $G$, the Young's modulus $E$ and Poisson's ratio $\nu$ can be obtained by $E=\frac{9 K G}{3 K+G}$ and $\nu=\frac{3 K-2 G}{2(3 K+G)}$, respectively.

\textcolor{black}{For 2D materials, VASPKIT assumes the crystal plane in the $xy$ plane. Then the relation between strain and stress can be written in the following form \cite{Zhang2017,Mazdziarz2019}}
\begin{equation}
\left(\begin{array}{c}{\sigma_{1}} \\ {\sigma_{2}} \\ {\sigma_{6}}\end{array}\right)=\left(\begin{array}{ccc}{\text{C}_{11}} & {\text{C}_{12}} & {\text{C}_{16}} \\ {\text{C}_{21}} & {\text{C}_{22}} & {\text{C}_{26}} \\ {\text{C}_{61}} & {\text{C}_{62}} & {\text{C}_{66}}\end{array}\right) \cdot\left(\begin{array}{c}{\varepsilon_{1}} \\ {\varepsilon_{2}} \\ {\varepsilon_{6}}\end{array}\right),
\end{equation}
where C$_{ij}$ (\emph{i},\emph{j}=1,2,6) is the in-plane stiffness tensor. The strain tensor $\boldsymbol{\epsilon}$ in Eq. (\ref{strain_tensor}) is simplified as

\begin{equation}
\boldsymbol{\epsilon}^{2D}=\left(\begin{array}{lll}
\varepsilon_{1} & \varepsilon_{6} / 2 & 0 \\
\varepsilon_{6} / 2 & \varepsilon_{2} & 0 \\
0 & 0 & 0
\end{array}\right).
\end{equation}

Then the elastic strain energy per unit area based on the strain-energy method can be expressed as \cite{Landau1959}

\begin{equation}
\begin{aligned}
\Delta E\left(S,\left\{\varepsilon_{i}\right\}\right)/S_0  =\frac{1}{2} (C_{11} \varepsilon_{1}^{2}+ C_{22} \varepsilon_{2}^{2}+2C_{12} \varepsilon_{1} \varepsilon_{2} \\ +  2C_{16} \varepsilon_{1} \varepsilon_{6}+2C_{26} \varepsilon_{2}\varepsilon_{6}+ C_{66} \varepsilon_{6}^{2}),
\end{aligned}
\end{equation}
where $S_{0}$ is the equilibrium area of the system. Clearly, the C$_{ij}$ is equal to the second partial derivative of strain energy $\Delta E$ with respect to strain $\varepsilon$, namely, $\text{C}_{ij}=(1/S_{0})(\partial^{2}\Delta E/\partial\varepsilon_{i}\partial\varepsilon_{j})$.  Therefore, the unit of elastic stiffness tensor for 2D materials is force per unit length (N/m). The classification of the different crystal system with the corresponding number of independent elastic constants and elastic stability conditions for 2D materials are summarized in Table \ref{iec_2d}.  \textcolor{black}{The details of strain modes and the derived elastic constants for each 2D crystal system based on energy-strain approach are listed in Appendix B.}
 
\begin{table*}[hbt]
\fontsize{9}{11}\selectfont
\setlength{\tabcolsep}{1.5pc}
\caption{Classification of crystal systems and independent elastic constants for 2D materials \cite{Mazdziarz2019}. In the last column, several prototype materials are shown.}
\label{iec_2d}
\begin{tabular*}{\textwidth}{@{}l@{\extracolsep{\fill}}cccc}
\hline
Crystal system &  Number of independent SOECs & Independent SOECs  & Material prototypes\\

\hline
Oblique & 6 & $\text{C}_{11}$, $\text{C}_{12}$, $\text{C}_{22}$, $\text{C}_{16}$, $\text{C}_{26}$, $\text{C}_{66}$  & - \\
Rectangle & 4 & $\text{C}_{11}$, $\text{C}_{12}$, $\text{C}_{22}$, $\text{C}_{66}$   & Borophene\\
Square & 3 & $\text{C}_{11}$,  $\text{C}_{12}$, $\text{C}_{66}$  & SnO \\
Hexagonal & 2 & $\text{C}_{11}$, $\text{C}_{12}$ & Graphene, MoS$_2$\\
\hline
\end{tabular*}
\end{table*}

\textcolor{black}{In order to provide a benchmark for computational studies}, we list the calculated second-order elastic constants for 2D and bulk prototype materials belonging to different crystal systems in Tables \ref{tec_3d} and \ref{tec_2d} respectively, together with other theoretical values \cite{Golesorkhtabar2013,Wei2009,Haastrup2018,Wang2015c} for \textcolor{black}{comparison} purposes. It is found that the results produced with different DFT codes are in good agreement with each other.

\begin{table*}[hbt]
\fontsize{8}{11}\selectfont
% space before first and after last column: 1.5pc
% space between columns: 3.0pc (twice the above)
\setlength{\tabcolsep}{0.5pc}
% -----------------------------------------------------
% adapted from TeX book, p. 241
\newlength{\digitwidth} \settowidth{\digitwidth}{\rm 0}
\catcode`?=\active \def?{\kern\digitwidth}
% -----------------------------------------------------
\caption{PBE-calculated elastic stiffness constants (in units of GPa) for ZrO$_2$, TiS$_2$, MgF$_2$, CaMoO$_4$, $\alpha$-Al$_2$O$_3$, CaMg(CO$_3$)$_2$, Ti and Diamond. For comparison purposes, the available theoretical values from the literature are also shown \cite{Golesorkhtabar2013}.}
\label{tec_3d}
\begin{tabular*}{\textwidth}{@{}l@{\extracolsep{\fill}}|ccccccccccccccccc}
\hline
                 & \multicolumn{2}{c}{ZrO$_2$.} 
                 & \multicolumn{2}{c}{TiS$_2$}    
                 & \multicolumn{2}{c}{MgF$_2$} 
                 & \multicolumn{2}{c}{CaMoO$_4$}    
                 & \multicolumn{2}{c}{$\alpha$-Al$_2$O$_3$} 
                 & \multicolumn{2}{c}{CaMg(CO$_3$)$_2$}  
                 & \multicolumn{2}{c}{Ti} 
                 & \multicolumn{2}{c}{Diamond}       \\           
\cline{2-3} \cline{4-5} \cline{6-7} \cline{8-9}  \cline{10-11} \cline{12-13}  \cline{14-15} \cline{16-17} 

C$_{ij}$ & Calc. & Ref. & Calc. & Ref. & Calc. & Ref. & Calc. & Ref. & Calc. & Ref. & Calc. & Ref. & Calc. & Ref. & Calc. & Ref.\\
\hline
C$_{11}$ &  334 & 334  &   314 &  312  &  134 &  130     & 130 & 126    & 452 & 451  & 192  & 194  & 184 &  189 & 1051 & 1052 \\
C$_{12}$ &  155 &  151  &  29 & 28  & 80 & 78   & 53 & 58     &  149 & 151  & 64&  67    & 83 & 85  & 127&  125 \\
C$_{13}$ &  82  & 82    &  78 & 84  & 59&  55   & 47 & 46     & 108 & 108   & 54&  57    & 78 & 74  &    &  \\
C$_{14}$ &    &        &     &      &    &       &     &        & 20 & 21     & 17 & 18    &         &   &   &  \\
C$_{15}$ & 26 & 32     &     &      &    &       &    &         &     &        & 13&  12    &         &   &   &  \\
C$_{16}$ &        &      &     &      & 10 & 10  &    &     &      &      &         &   &   &   &   &  \\
C$_{22}$ &  352&  356  &311&  306  &   &        &    &         &     &        &    &        &         &   &   &  \\
C$_{23}$ & 146&  142   &  25&  21  &    &       &    &         &     &        &    &        &         &   &   &  \\
C$_{24}$ &            &      &     &     &      &     &        &      &       &     &       &         &   &   &   &   \\
C$_{25}$ &  5&   2     &     &      &    &       &    &         &     &        &    &        &         &   &   &  \\
C$_{26}$ &    &        &     &      &    &       &    &         &     &        &     &       &         &   &   &  \\
C$_{33}$ & 263 & 251   &404 & 406  &192 & 185  & 112&  110   & 455 & 452   & 107 & 108      & 197  & 187   &   &  \\ 
C$_{34}$ &            &     &      &    &       &    &         &      &       &     &       &         &   &   &   &  \\
C$_{35}$ & 2 & 7       &    &       &    &       &    &         &      &       &    &        &         &   &   &  \\
C$_{36}$ &            &     &      &     &      &     &        &      &      &     &       &         &   &   &    &   \\
C$_{44}$ & 78 & 71     & 73 &  73   & 52 & 61   & 30&  29     & 133&  132   & 37 & 39    & 46&  41  & 560&  559 \\
C$_{45}$ &            &      &      &    &       &    &         &      &       &    &        &         &   &   &   &  \\
C$_{46}$ &  15 &   15   &    &       &    &       &   &          &     &        &   &         &         &   &   &  \\
C$_{55}$ & 70 &  71     &100  & 106  &   &        &   &          &    &         &    &        &         &  &   &  \\
C$_{56}$ &    &        &     &      &    &       &    &         &      &       &     &       &         &   &   &  \\
C$_{66}$ & 113&  115   &118 &  117  & 90 & 83   & 38&  34     &     &        &    &        &         &   &   &  \\
\hline
\end{tabular*}
\end{table*}

\begin{table*}[htbp]
\fontsize{9}{11}\selectfont
\caption{\label{tec_2d} PBE-calculated in-plane elastic stiffness constants (in units of N/m). For comparison purposes, the available theoretical or experimental values from the previous literature are also shown.} 
\begin{tabular}{ccccccccc}
\hline
&\multicolumn{2}{c}{C$_{11}$}
&\multicolumn{2}{c}{C$_{22}$}
&\multicolumn{2}{c}{C$_{12}$ }
&\multicolumn{2}{c}{C$_{66}$}\\
Systems & Our work &  Literature & Our work & Literature  & Our work & Literature  & Our work & Literature     \\
\hline
Graphene & 349.1    & 358.1  \cite{Wei2009}  &    &   & 60.3 &  60.4 \cite{Wei2009}  &    &      \\
MoS$_2$  & 128.9    & 131.4 \cite{Haastrup2018} &  &   & 32.6   &  32.6 \cite{Haastrup2018} &    &   \\
SnO &  48.14    &    &    &   & 38.9 &    & 39.0  &      \\
Phosphorene & 104.4    & 105.2  \cite{Wang2015c}  & 34.0  & 26.2 \cite{Wang2015c} & 21.6 & 18.4 \cite{Wang2015c}   & 27.4  &      \\
\hline
\end{tabular} 

\end{table*}

\subsection{Equations of State}
\textcolor{black}{Thermodynamic equations of state (EOS) for crystalline solids describe the relationships among the internal energy $E$, pressure $P$, volume $V$ and temperature $T$. It plays a crucial role in  predicting the structural and thermodynamical properties of materials under high pressure and high temperature in condensed matter sciences \cite{Latimer2018}, especially in extreme conditions such as earth or planetary interiors where the properties of materials are quite different from those found at ambient conditions \cite{Anderson1995}. Various EOS formulas have been proposed. One of the most widely used isothermal EOSs in solid state physics is Murnaghan EOS model assuming that the bulk modulus varies linearly with pressure \cite{Murnaghan1944}. The resulting energy–volume relationship is given as:}

\begin{equation}
E(\nu)=E_{0}+\frac{B V_{0}}{(C+1)}\left(\frac{\nu^{-C}-1}{C}+\nu-1\right),
\end{equation}
\textcolor{black}{where $\nu=\frac{V}{V_{0}}$, $V_{0}$ and $E_0$ are the volume and energy at zero pressure respectively. The values of bulk modulus $K$ and its pressure derivative $K^{\prime}$ can be further deduced in terms of the fitting parameters $B$ and $C$. The bulk modulus $K$ is a measure of the resistance of a solid material to compression. It is defined as the proportion of volumetric stress related to the volumetric strain for any material, namely,}

\begin{equation}
K=-V\left(\frac{\partial P}{\partial V}\right)_{T}.
\end{equation}

\begin{figure}[htbp]
\centering
\includegraphics[scale=0.45]{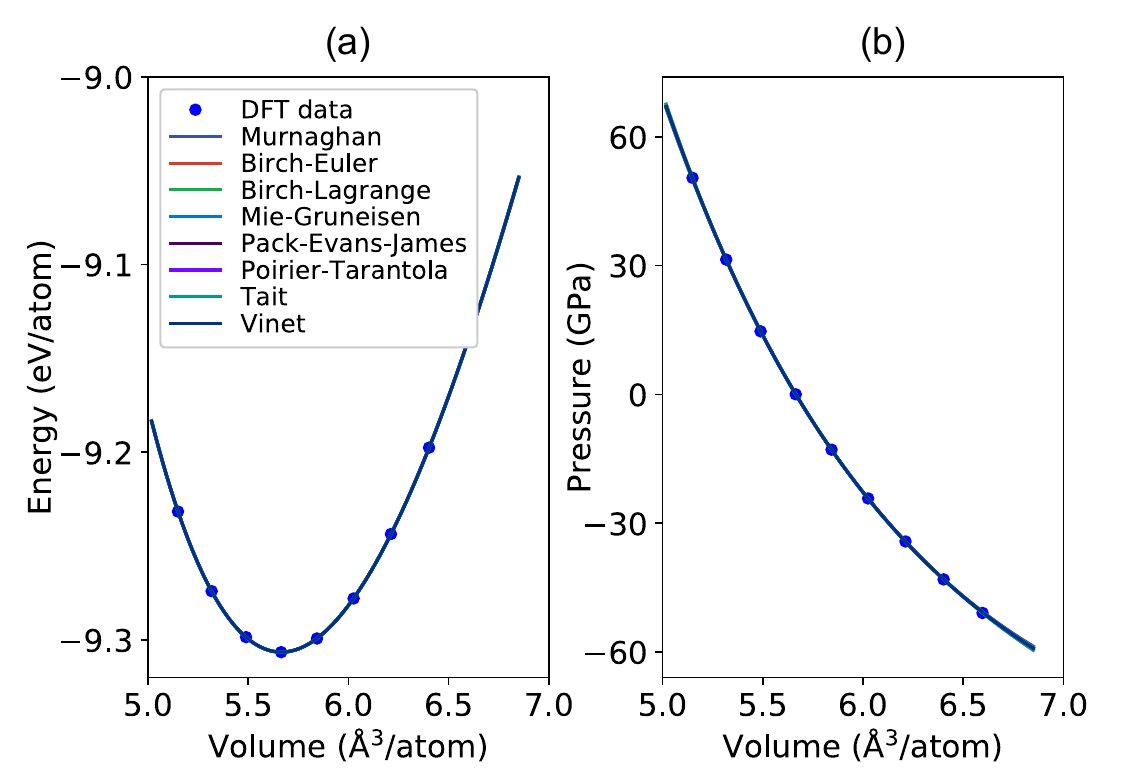}
\caption{\label{fig_eos}(Color online) The equations of states of diamond using different EOS models as listed in Table \ref{teos}.}
\end{figure}

\textcolor{black}{The workflow of EOS utility is similar to that of the elastic constants presented in Fig. \ref{elastic}. In addition to the equilibrium volume and bulk modulus, pressure and energy as functions of volume are also provided in this utility. Very recently, Latimer $\emph{et al}$. evaluated the quality of fit for the 8 widely-used EOS models listed in Table \ref{teos} across 87 elements and over 100 compounds \cite{{Latimer2018}}. They pointed out that it is hard to find a universal EOS model applicable to all types of solids and accurate over the whole range of pressure.
Furthermore,  their results reveal that the Birch-Euler \cite{Birch1947}, Tait \cite{Dymond1988}, and Vinet \cite{Vinet1987} models give the best overall quality of fit to \textcolor{black}{the calculated energy-volume curves among all the} equations under examination. However, the \textcolor{black}{inconsistencies} among these investigated equations is not significant.  As a benchmark test, \textcolor{black}{the calculated energy and pressure of diamond as a function of volume using different EOS models are presented in Fig. \ref{fig_eos}.} One can clearly find that the agreement among these EOS fits is very satisfactory on the whole. The calculated bulk modulus ranges from 440 GPa to 442 GPa, in good agreement with the experimental value of 443 GPa \cite{Hebbache2004}}.

\begin{table*}[htbp]
\centering
%\fontsize{9}{11}\selectfont
%\begin{ruledtabular}
\caption{\label{teos} The analytic formulae of energy-volume relation and bulk modulus $K$ for several widely used EOS models based on Table 1 of Ref. \cite{Latimer2018}.}
\begin{tabular}{ccc}
\hline
Model & Internal energy $E$ & Bulk modulus $K$ ($\nu=1$)   \\
\hline
Birch (Euler) \cite{Birch1947} &  $E=E_{0}+B V_{0}\left(\left(\nu^{-\frac{2}{3}}-1\right)^{2}+\frac{c}{2}\left(\nu^{-\frac{2}{3}}-1\right)^{3}\right)$ & $\frac{8 B}{9}$     \\
Birch (Lagrange) \cite{Birch1947}  & $E=E_{0}+B V_{0} C-B V_{o} \nu^{\frac{2}{3}}\left((C-2)\left(1-\nu^{\frac{2}{3}}\right)^{2}+C\left(1-\nu^{\frac{2}{3}}\right)+C\right)$ & $\frac{16 B}{9}$ \\
Mie-Gruneisen \cite{Roy2005} &  $E=E_{0}+\frac{B V_{0}}{C}-\frac{B V_{0}}{C-1}\left(\nu^{-\frac{1}{3}}-\frac{1}{C} \nu^{-\frac{c}{3}}\right)$ & $\frac{B}{9}$  \\
Murnaghan \cite{Murnaghan1944} & $E=E_{0}+\frac{B V_{0}}{(C+1)}\left(\frac{\nu^{-c}-1}{C}+\nu-1\right)$ & $B$    \\
Pack-Evans-James  \cite{Pack1948}  & $E=E_{0}+\frac{B V_{0}}{c}\left(\frac{1}{c}\left(e^{3 C\left(1-\nu^{\frac{1}{3}}\right)}-1\right)-3\left(1-\nu^{\frac{1}{3}}\right)\right)$ & $B$  \\
Poirier-Tarantola \cite{Poirier1998} & $E=E_{0}+B V_{0}(\ln (\nu))^{2}(3-C(\ln (\nu)))$ &  $6B$    \\
Tait \cite{Dymond1988} & $E=E_{0}+\frac{B V_{0}}{C}\left(\nu-1+\frac{1}{C}\left(e^{C(1-\nu)}-1\right)\right)$ & $B$   \\
Vinet \cite{Vinet1987} & $E=E_{0}+\frac{B V_{0}}{C^{2}}\left(1-\left(1+C\left(\nu^{\frac{1}{2}}-1\right)\right) e^{-C\left(\nu^{\frac{1}{3}}-1\right)}\right)$ & $\frac{B}{9}$  \\
\hline
\end{tabular}

\end{table*}

\subsection{Band Structure and Density of States}
\textcolor{black}{Band structure is one of the most important concepts in solid state physics. It provides the electronic levels in crystal structures, which are characterized by two quantum numbers, the band index $n$ and the Bloch vector $\mathbf{k}$ along high symmetry directions in the BZ. Besides the band structure, the density of states (DOS) is another quantity that is defined as the number of states per interval of energy at each energy level that are available to be occupied by electrons. A high DOS at a specific energy level means that there are many states available for occupation and zero DOS means that no state can be occupied at that energy level. DOS can be used to calculate the density of free charge carriers in semiconductors, the electronic contribution to the heat capacity in metals. Moreover, it also provides an indirect description for properties such as magnetism, chemical bonding, optical absorption spectrum, and etc.}

\begin{figure*}[htbp]
\centering
\includegraphics[scale=0.5]{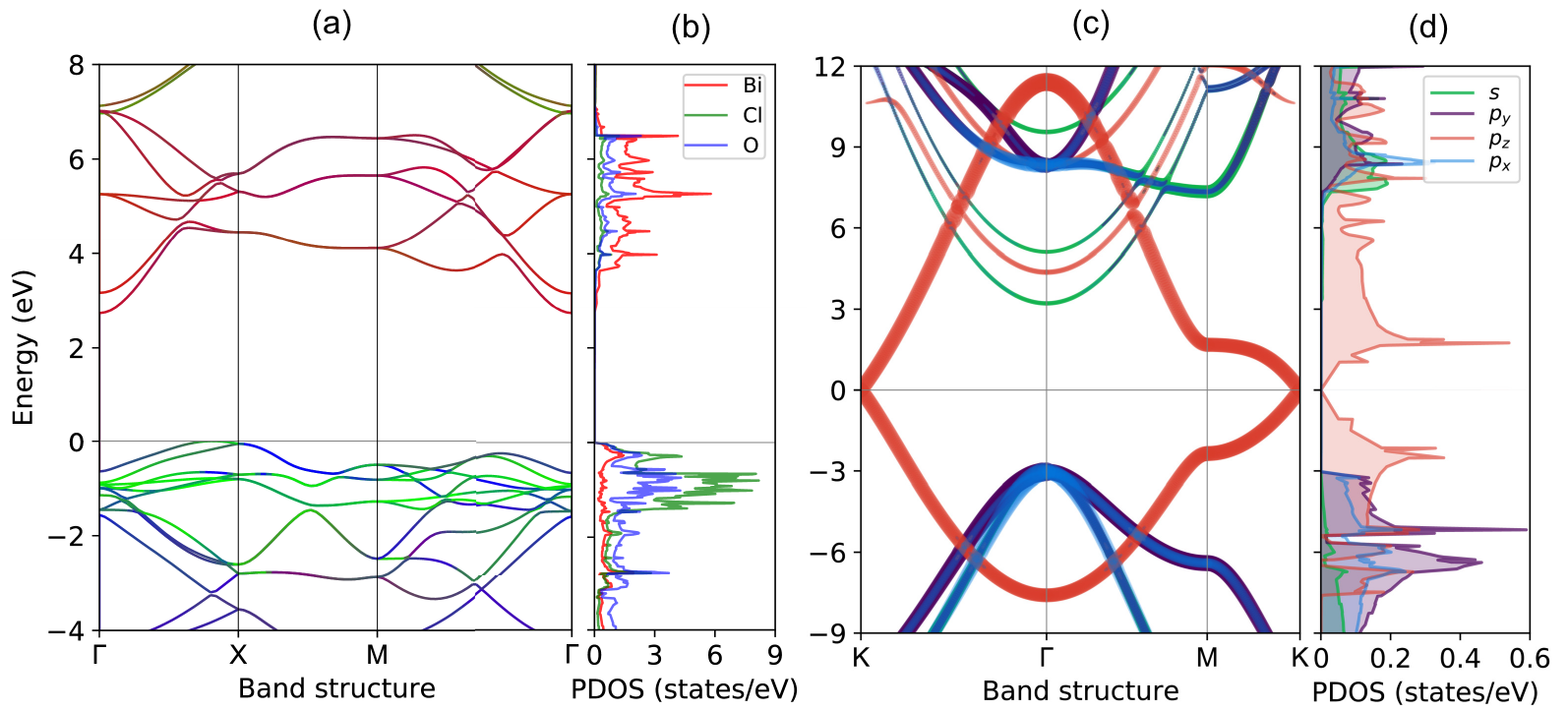}
\caption{\label{pband}(Color online) Projected band structure (left panel) and density of states (right panel) of (a) BiClO ($P4/nmm$) and (b) graphene monolayers. The Fermi energy is set to zero eV. }
\end{figure*}

\begin{figure*}[htbp]
\centering
\includegraphics[scale=0.22]{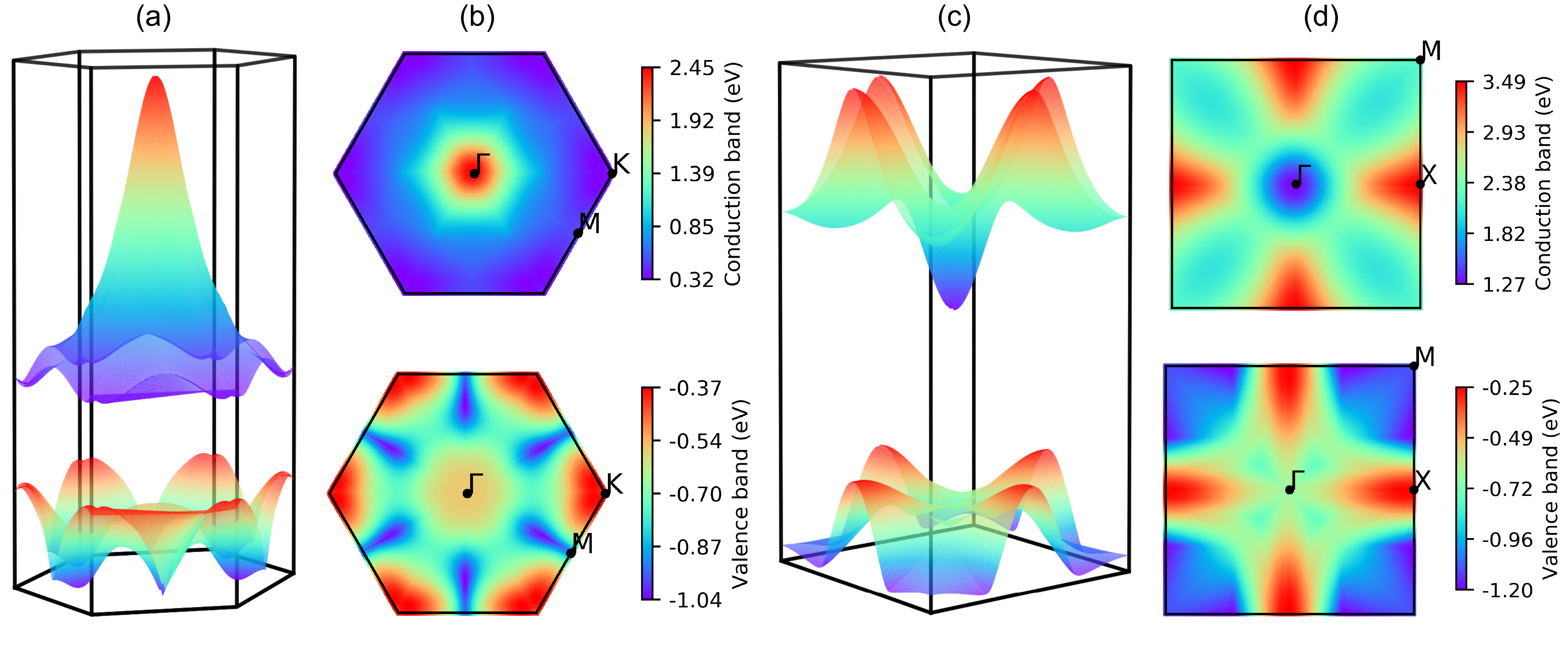}
\caption{\label{3d_band}(Color online) The global band structures of the highest valence and lowest conduction bands  for (a) MoTe$_2$ ($P\overline{6}m2$) and (b) BiIO ($P4/nmm$) monolayers. The Fermi energy is set to zero.}
\end{figure*}

In addition to the conventional plain band structure, VASPKIT can also deal with the projected band which provides insight into the atomic orbital contributions in each state. As illustrated examples, the projected band structures and density of states (DOS) of BiClO ($P4/nmm$)  and graphene monolayers are depicted in Fig. \ref{pband}. To illustrate the band dispersion anisotropy of 2D materials, the 3D global band structures of the highest valence and lowest conduction bands for MoTe$_2$ ($P\overline{6}m2$) and BiIO are shown in Fig. \ref{3d_band}.  

\subsection{Effective Masses of Carriers}
Generally,  the band dispersions close to conduction or valence band extrema can be approximated as parabolic for the semiconductors with low carrier concentrations. Consequently, the analytical expression of effective masses of carriers (EMC) $m^{*}$ for electrons and holes (in units of electron mass $m_{0}$) is given by

\begin{equation}
m^{*}=\hbar^{2}\left[\frac{\partial^{2} E(k)}{\partial^{2} k}\right]^{-1},
\end{equation}
where $E(k)$ are the energy dispersion relation functions described by band structures, and $\hbar$ is the reduced Planck constant. Clearly, $m^{*}$ is inversely proportional to the curvature of the electronic dispersion in reciprocal space, implying that CB and VB edges with larger dispersions result in smaller effective masses. It is noteworthy that the \textcolor{black}{above} expression should not be used in non-parabolic band dispersion cases, for example, the linear dispersion in the band edges of graphene \cite{Whalley2019}. Similarly, the Fermi velocity represents the group velocity of electrons traveling in the material \textcolor{black}{is} defined as 

\begin{equation}
v_{F}=\frac{1}{\hbar} \frac{\partial E}{\partial k}.
\end{equation} 

\begin{figure}[htbp]
\centering
\includegraphics[scale=0.5]{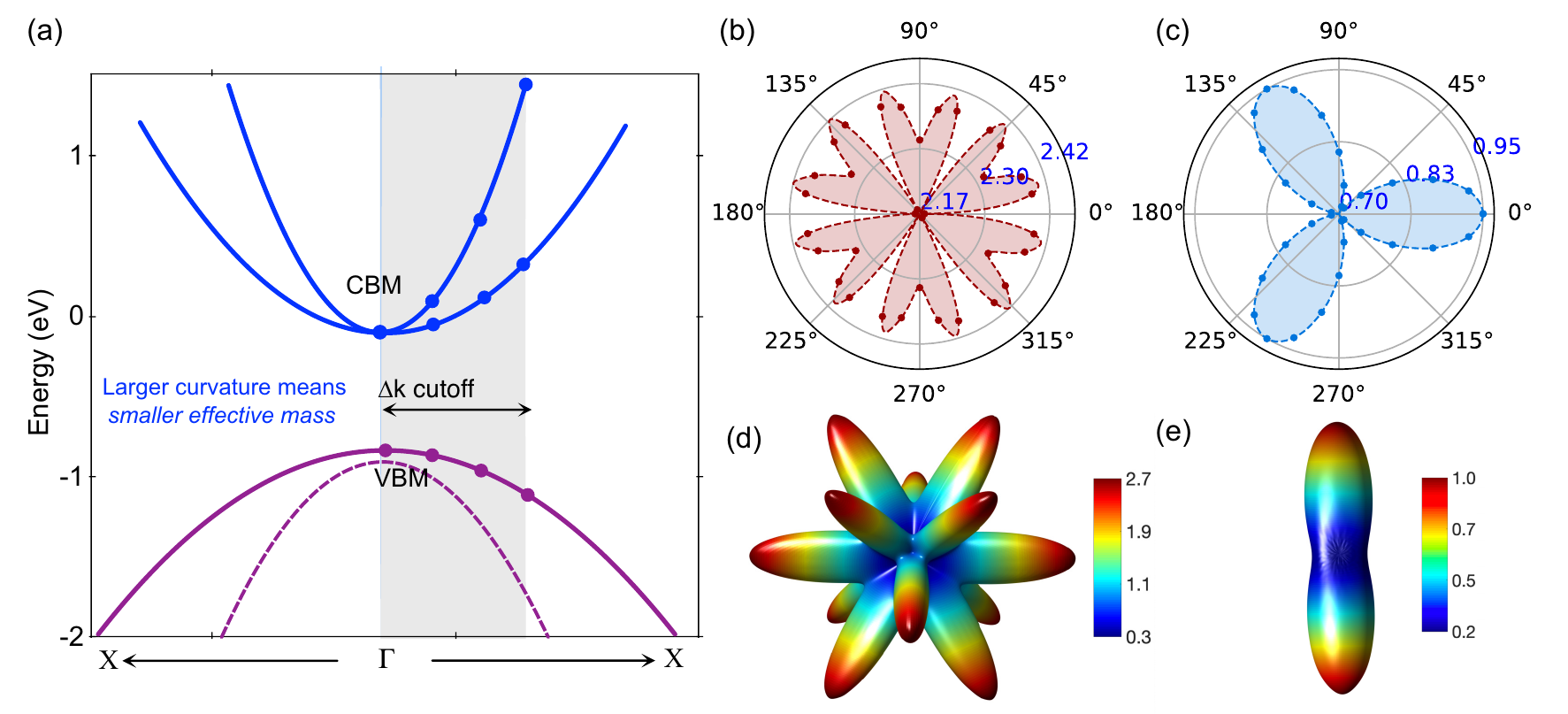}
\caption{\label{emc_anisotropy}(Color online) (a) Schematic illustration of the determination of effective masses based on second-order polynomial fitting  around the conduction and valence band extrema. Orientation-dependent effective masses (in units of electron mass $m_0$) of (b, d) hole and (c, e) electron carriers for 2D BN monolayer (b, c) and bulk Si (d, e)  respectively. }
\end{figure}

Figure \ref{emc_anisotropy} (a) illustrates schematically the determination of effective masses by fitting the band dispersion \textcolor{black}{with} a second order polynomial. The effective masses of carriers are calculated \textcolor{black}{using} an ultrafine $k$-mesh of density uniformly distributed inside a circle of radius $k$-cutoff. Haastrup \emph{et al}. pointed out that the inclusion of third order terms stabilizes the fitting procedure and yields the effective masses that are less sensitive to the details of the employed $k$-mesh \cite{Haastrup2018}. Thus, a third order polynomial is also adopted to fit the band energy curvature in the EMC utility. In Table \ref{tab_emc} we show the calculated effective masses for several typical 2D and bulk semiconductors with available effective mass data, including Phosphorene \cite{Haastrup2018}, MoS$_2$ \cite{Haastrup2018}, GaAs \cite{Williamson2000} and Diamond \cite{Naka2013}. Overall, the agreement is very \textcolor{black}{good}. In addition, the EMC utility can also calculate the orientation-dependent effective masses of charge carriers. Examples for this functionality are shown in Figs. \ref{emc_anisotropy} (b)-(e).  One can find that the calculated effective masses of two investigated systems show strong anisotropy, especially for the case of bulk Si. 

\begin{table*}
\fontsize{8}{11}\selectfont
\caption{\label{tab_emc} The calculated effective masses of electron $m_e$ and hole $m_h$ carriers (in units of the electron mass m$_0$) for typical semiconductors using PBE approach. The masses are labeled by the band extremum and the direction of the hight symmetry line along which the mass is calculated using a simple parabolic line fit. The labels of high-symmetry points are adopted from the Ref. \cite{Hinuma2017}.} 
\begin{tabular}{cccccc}
\hline
&\multicolumn{1}{c}{}
&\multicolumn{2}{c}{Electron mass ($m_e$)}
&\multicolumn{2}{c}{Hole mass ($m_h$)}\\
%\hline
Material & Direction & Our work &  Literature & Our work & Literature     \\
\hline
Phosphorene  & $\Gamma$$\rightarrow$$\text{X}$ (zig-zag) & 1.23 & 1.24 \cite{Haastrup2018} & 7.21  & 6.56  \cite{Haastrup2018} \\
Phosphorene & $\Gamma$$\rightarrow$$\text{Y}$ (armchair) & 0.19 & 0.14 \cite{Haastrup2018} & 0.17  & 0.13  \cite{Haastrup2018} \\
MoS$_2$ monolayer & $\text{K}$$\rightarrow$$\Gamma$ &0.47 & 0.42 \cite{Haastrup2018} &0.56 & 0.53 \cite{Haastrup2018} \\
GaAs bulk & $\Gamma$$\rightarrow$$\text{X}$ &0.06 & 0.07 \cite{Williamson2000} &0.35  & 0.34 \cite{Williamson2000} \\
Diamond bulk & $\Gamma$$\rightarrow$$\text{X}$  &0.32 & 0.29 \cite{Naka2013}  &0.27 & 0.36 \cite{Naka2013}  \\
\hline
\end{tabular} 
\end{table*}

\subsection{Charge Density and Potential Manipulation}

For spin-polarized systems, the charge density $\rho(\mathbf{r})$ and magnetization (spin) density $m(\mathbf{r})$ are defined as 

\begin{equation}
\begin{aligned}
\rho(\mathbf{r})=\rho_{\uparrow}(\mathbf{r})+\rho_{\downarrow}(\mathbf{r}) \\
m(\mathbf{r})=\rho_{\uparrow}(\mathbf{r})-\rho_{\downarrow}(\mathbf{r})
\end{aligned},
\end{equation}
where $\rho_{\uparrow}(\mathbf{r})$ and $\rho_{\downarrow}(\mathbf{r})$ are the spin-up and spin-down \textcolor{black}{densities. Note} that the $\rho_{\uparrow}(\mathbf{r}) = \rho_{\downarrow}(\mathbf{r})$ in non-spin-polarized cases. The spin density $\rho_{\sigma}(\mathbf{r})$ is expressed as

\begin{equation}
\rho_{\sigma}(\mathbf{r})=\sum_{o c c} \varphi_{i \sigma}^{*}(\mathbf{r}) \varphi_{i \sigma}(\mathbf{r}),
\end{equation}
where ${\sigma}$ and $i$ are the spin- and band-index respectively, $\varphi_{i \sigma}(\mathbf{r})$ is the normalized single-particle wave-function. $occ$ means that \textcolor{black}{summation is} over all occupied states. 

The charge density difference $\Delta \rho(\mathbf{r})$ can track the charge transfer \textcolor{black}{and gain information of the interaction between the two parts that constitute the system.} The $\Delta \rho_(\mathbf{r})$ can be obtained

\begin{equation}
\Delta \rho(\mathbf{r})=\rho_{\mathrm{AB}}(\mathbf{r})-\rho_{\mathrm{A}}(\mathbf{r})-\rho_{\mathrm{B}}(\mathbf{r}),
\end{equation}
where $\rho_{\mathrm{A}}(\mathbf{r})$, $\rho_{\mathrm{B}}(\mathbf{r})$ and $\rho_{\mathrm{AB}}(\mathbf{r})$ are the  charge density of  reactants A and B, and product C. VAPSKIT can extract charge-density, spin-density, electrostatic potential as well as the difference of these quantities, and save \textcolor{black}{them in}  VESTA (.vasp) \cite{Kresse1996,Kresse1996a,Momma2011}, XCrysDen (.xsf) \cite{Kokalj2003}, or Gaussian (.cube) formats \cite{Frisch2009}. 

From the three-dimensional electronic charge density and electrostatic potential one can get the average one-dimensional charge density $\overline{n}(z)$ and electrostatic potential $
\overline{V}(z)$ by calculating the planar average function $(\overline{f})$ \cite{Peressi1998}:

\begin{figure}[htbp]
\centering
\includegraphics[scale=0.19]{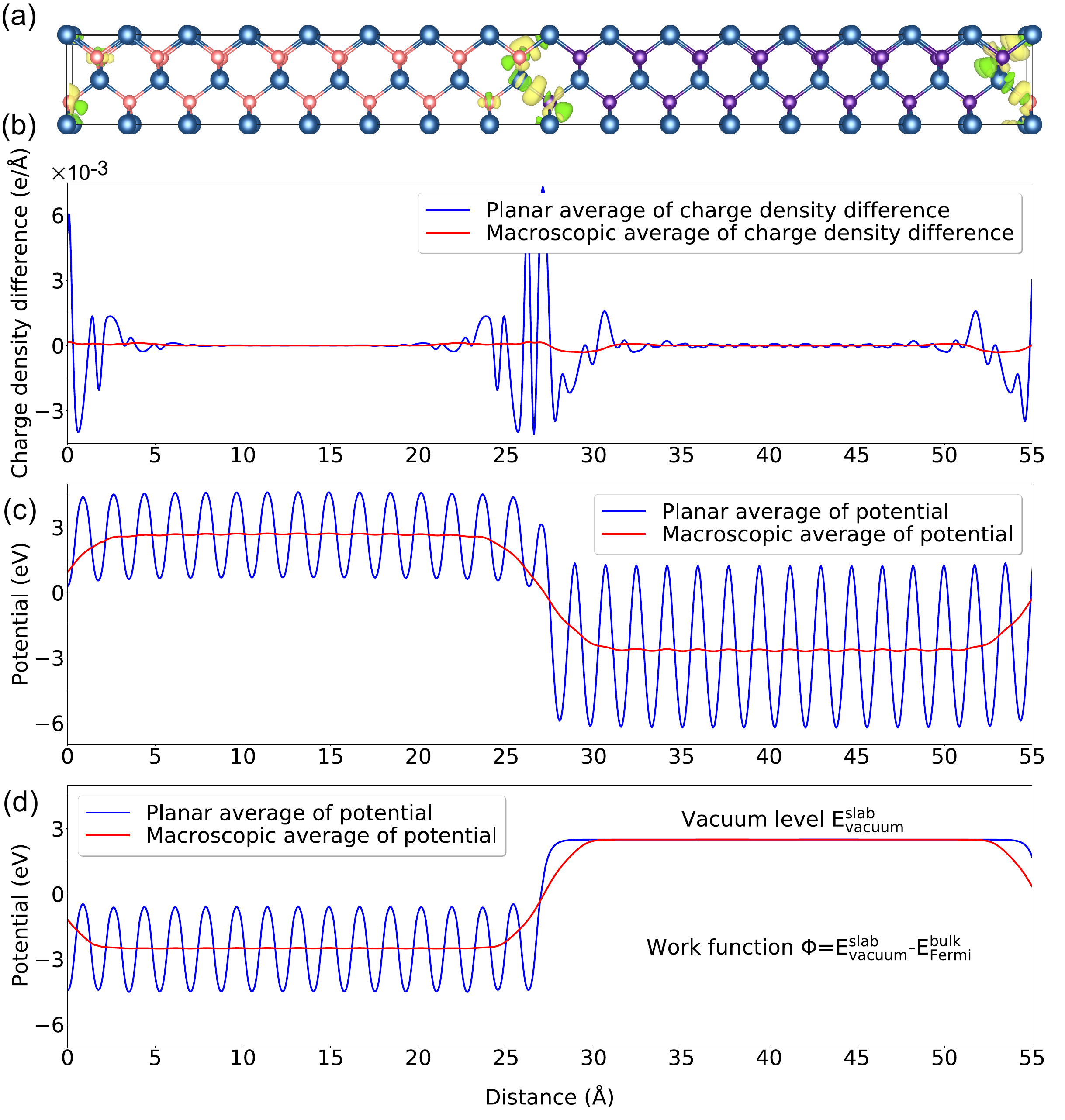}
\caption{\label{macro}(Color online) Calculated (a) charge density difference, planar- (blue line) and macroscopic averages (red line) of (b) charge density difference, (c) electrostatic potential of a GaAs/AlAs (100) heterojunction, and (d) electrostatic potential of a GaAs (110) slab. Ga atoms are shown in purple, As are blue, and Al are red. }
\end{figure}

\begin{equation}
\overline{f}(z)=\frac{1}{S} \int_{S} V(\mathbf{r})dxdy,
\end{equation}
where $S$ represents the area of a unit cell in the $x-y$ plane. Generally, this planar-averaged charge density and potential exhibit periodic oscillations along the $z$ axis due to the spatial distribution of the electrons and ionic cores. These oscillations can be removed using a macroscopic averaging procedure \cite{Peressi1998}:

\begin{equation}
\overline{\overline{f}}(z)=\frac{1}{L} \int_{-L / 2}^{L / 2} \overline{f}(z) dz,
\end{equation}
where $L$ is the length of the period of oscillation along $z$. By definition, this macroscopic average would produce a constant value in the bulk. It is expected to reach a plateau value in the bulk-like regions of each layer in the superlattice. As an example, Figure \ref{macro} shows the calculated planar and macroscopic averages of charge density difference and electrostatic potential for a (100)-oriented GaAs/AlAs heterojunction and a (110)-oriented GaAs slab, respectively. 

\subsection{Fermi surface}

\begin{figure}[htbp]
\centering
\includegraphics[scale=0.6]{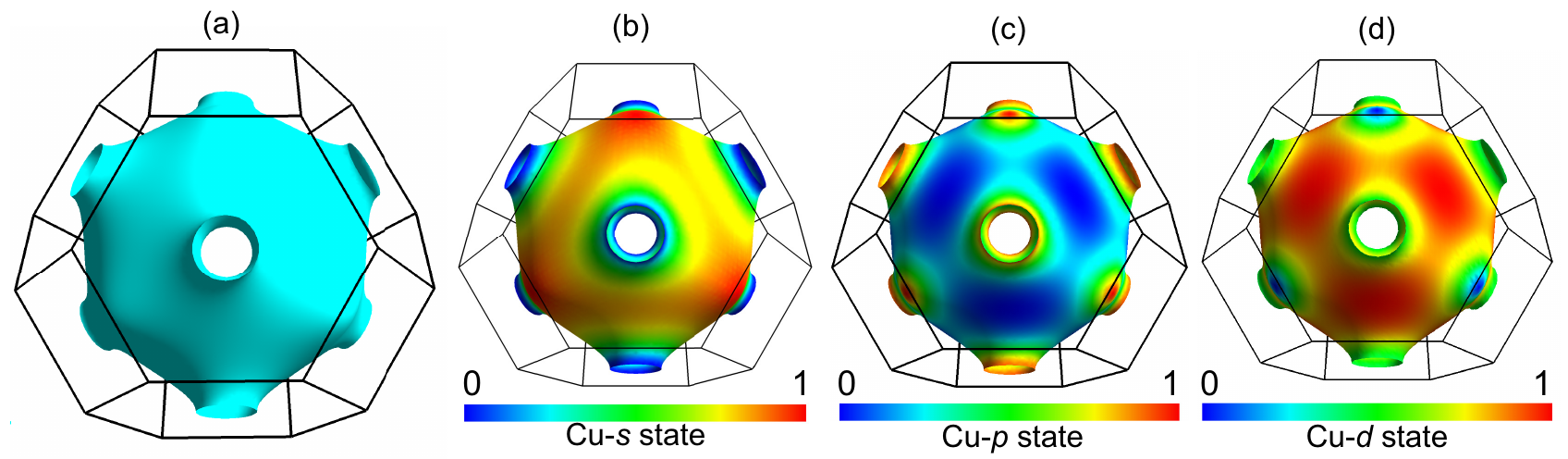}
\caption{\label{FermiSurface}(Color online) (a) Plain Fermi surface of Cu.  Orbital-resolved fermi surface of (b) Cu-$s$, (c) Cu-$p$  and (d) Cu-$d$ states respectively, visualized by the FermiSurfser package \cite{Kawamura2019}. The color denotes the weight of the states.}
\end{figure}

\textcolor{black}{Fermi surface is the surface in reciprocal space which separates occupied from unoccupied electron states at zero temperature \cite{Dugdale2016}. It is defined to be the set of $k$-points such that $E(\mathbf{k})=\mu$ for any band index $n$, where $\mu$ is the Fermi energy. The shape of the Fermi surface is derived from the periodicity and symmetry of the crystalline lattice as well as the occupation of electronic energy bands. 
The knowledge of the topology of the Fermi surface is important for characterizing and predicting the thermal, electronic and magnetic properties. To calculate the Fermi surface, one first needs to use VASPKIT to determine the $k$-mesh $N_1{\times}N_2{\times}N_3$ based on the specified $k$-spacing value. The $k$-spacing is defined as the smallest allowed spacing between the $k$-points in BZ, that is, $N_i=\max \left(1,\left|\mathbf{b}_i\right|/k\mathrm{spacing}\right)$, where $\left|\mathbf{b}_i\right|$ is the length of the reciprocal lattice vector in the $i-th$ direction. To reduce the computational cost, only the eigenvalues at the inequivalent $k$-points in irreducible Brillouin zone are calculated using VASP. Then these $k$-points with the sum of the corresponding weight can be mapped to fill the entire BZ using symmetry operations without approximation during the post-processing. The resulting Fermi surface can be visualized using the XcrysDen \cite{Kokalj2003} or FermiSurfser programs \cite{Kawamura2019}. To illustrate the capabilities of this utility,} the calculated Fermi surfaces of copper colored by the atomic orbital projected-weights are shown in Fig. \ref{FermiSurface}.

\subsection{Wave-Function Visualization}

\begin{figure}[htbp]
\centering
\includegraphics[scale=0.44]{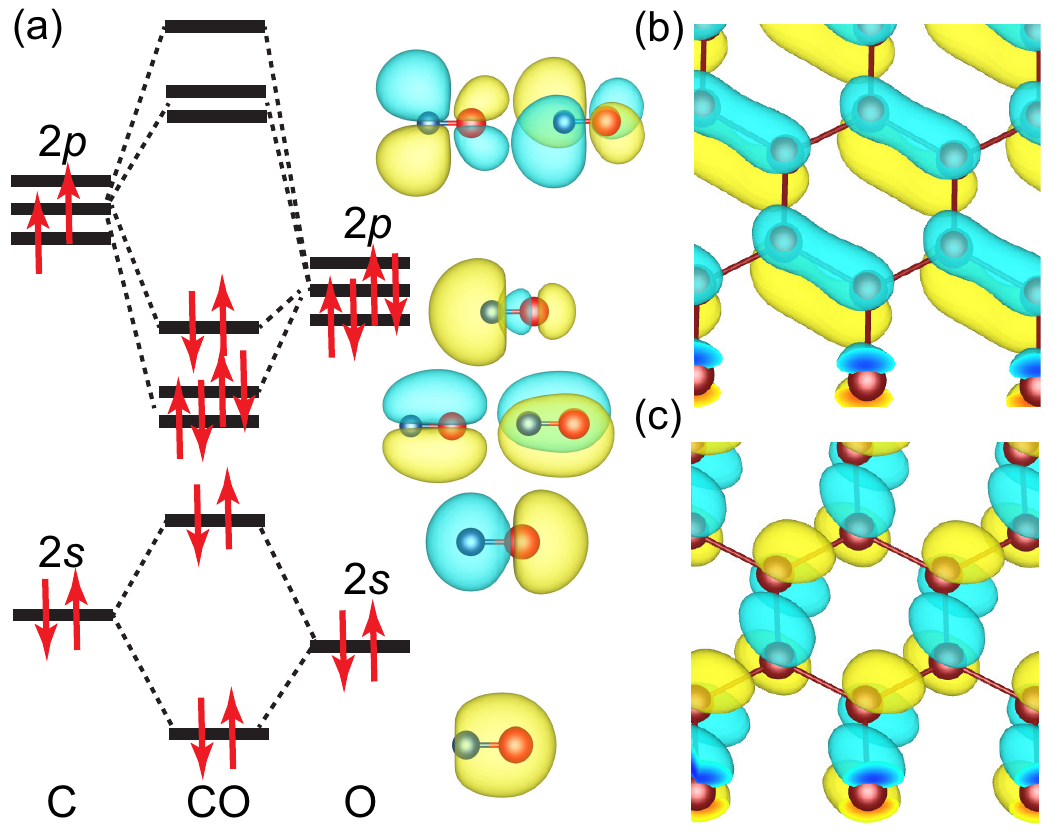}
\caption{\label{MO}(Color online) Calculated isosurfaces of wave functions in real space for (a) CO molecule, (b) VBM and (c) CBM for graphene respectively, visualized by the VESTA package \cite{Momma2011}.}
\end{figure}

To visualize wave functions, VASPKIT first reads the plane wave (PW) coefficients $\psi_{m\mathbf{k}}(\mathbf{k})$  of the specified wave-vector $\mathbf{k}$ point and band-index $m$ from the WAVECAR file, and performs a fast Fourier transform algorithm to convert the $\psi_{m\mathbf{k}}(\mathbf{k})$ from the reciprocal space to the real space, as denoted by $\psi_{m\mathbf{k}}(\mathbf{r})$. The $\psi_{m\mathbf{k}}(\mathbf{r})$ can thus be obtained

\begin{equation}
\label{pw}
\psi_{m\mathbf{k}}(\mathbf{r})= \sum_{\mathbf{G}} C{_{m\mathbf{k}}({\mathbf{k}+\mathbf{G})}} \mathrm{e}^{\mathrm{i}(\mathbf{k}+\mathbf{G}) \cdot \mathbf{r}},
\end{equation}
\textcolor{black}{where $\mathbf{G}$ is the reciprocal lattice vector, and $C{_{m\mathbf{k}}({\mathbf{k}+\mathbf{G})}}$ is the plane-wave coefficient of the wave vector $\mathbf{k+G}$ and band-index $m$ in reciprocal space}. Examples of the calculated wave function plots in real space are shown in Fig. \ref{MO}.

\subsection{Band Structure Unfolding}
The electronic structures of real materials are perturbed by structural defects, impurities, fluctuations of the chemical composition, and etc.  In DFT calculations, these defects and incommensurate structures are usually investigated by using SC models. \textcolor{black}{Nevertheless, it is difficult to compare directly the SC band structure with the PC band structure due to the folding of the bands into the smaller SC Brillouin zone (SBZ).}  Popescu and Zunger proposed the effective band structures (EBS) method which can unfold the SC band structures  into the corresponding PC Brillouin zone (pbz) \cite{Popescu2010,Popescu2012}. Such a delicate technique greatly simplifies the analysis of the results and enable direct comparisons with electronic structures of pristine materials.

As aforementioned, the lattice vectors of the SC and PC satisfy $\mathbf{A}=\mathbf{M} \cdot \mathbf{a}$, where $\mathbf{A}$ and $\mathbf{a}$ are the lattice vectors of SC and PC. The elements of transformation matrix $\mathbf{M}$ are integers $\left(m_{i j} \in \mathbb{Z}\right)$ when building SC from PC.  \textcolor{black}{In the band unfolding utility, the transformation matrix is not required to be diagonal. In other words, the SC and PC lattice vectors do not need to be collinear. Following} a general convention, capital and lower case letters indicate the quantities in the SC and PC respectively unless otherwise stated. A similar relation holds in reciprocal space: 

\begin{equation}
\mathbf{B}=\left({\mathbf{M}}^{-1}\right)^{T} \cdot \mathbf{b},
\end{equation}

where $\mathbf{B}$ and $\mathbf{b}$  are  the reciprocal lattice vectors of the SC and PC respectively. The reciprocal lattice vectors $\mathbf{g}_{n}\left(\mathbf{G}_{m}\right)$ in the pbz (SBZ) \textcolor{black}{are} expressed as
$$
\begin{array}{l}{\mathbf{g}_{n}=\sum_{i} n_{i} \mathbf{b}_{i}, \quad n_{i} \in \mathbb{Z}} \\ {\mathbf{G}_{m}=\sum_{i} m_{i} \mathbf{B}_{i}, \quad m_{i} \in \mathbb{Z}}\end{array},
$$
where $\left\{\mathbf{g}_{n}\right\} \subset\left\{\mathbf{G}_{m}\right\}$, i.e., every reciprocal lattice vector of the pbz is also one of the SBZ.

For a given $\mathbf{k}$ in pbz, there is a $\mathbf{K}$ in the SBZ to which it folds into, and the two vectors are related by a reciprocal lattice vector $\mathbf{G}$ in the SBZ:

\begin{equation}
\mathbf{k}=\mathbf{K}+\mathbf{G}_i, i=1, \ldots, N_{\mathbf{K}},
\end{equation}
where $N_{\mathbf{K}}$ is the determinant $|M|$ that determine the the multiplicity of the SC. When choosing plane waves as basis functions, \textcolor{black}{The projection of the SC eigenstates $|\psi_{m\mathbf{K}}^{\mathrm{SC}}\rangle$ on the PC eigenstates $|\psi_{n\mathbf{k}}^{\mathrm{PC}}\rangle$  is given by the spectra weight $P_{\mathbf{K} m}$} \cite{Popescu2010,Popescu2012}:

\begin{equation}\label{ebs}
\begin{aligned} P_{\mathbf{K} m}\left(\mathbf{k}_{i}\right) &=\sum_{n}\left|\left\langle\psi_{m\mathbf{K}}^{\mathrm{SC}} \mid \psi_{n \mathbf{k}}^{\mathrm{PC}}\right\rangle\right|^{2}=\sum_{\mathbf{g}}\left|C_{m\mathbf{K} }\left(\mathbf{g}+\mathbf{k}_{i}-\mathbf{K}\right)\right|^{2} \\ &=\sum_{\mathbf{g}}\left|C_{ m\mathbf{K}}\left(\mathbf{g}+\mathbf{G}_{i}\right)\right|^{2} \end{aligned}, 
\end{equation}
\textcolor{black}{where $m$ and $n$ stand for band indices at vectors $\mathbf{K}$ and $\mathbf{k}_{i}$ in the reciprocal space of the SC and PC, respectively. $C_{ m\mathbf{K}}$ is the PW coefficients given by Eq. (\ref{pw}) that span the eigenstates of the SC. This implies that the required information about the PC is the reciprocal lattice vectors of the primitive cell $\mathbf{g}$ only and the knowledge of the PC eigenstates is not necessary. Clearly, all the filtered $C_{m\mathbf{K}}\left(\mathbf{g}+\mathbf{G}_{j}\right)$ coefficients only contribute to the spectral function. The quantity $P_{\mathbf{K} m}$ represents the amount of Bloch character $\mathbf{k}_{i}$ preserved in $|\psi_{n\mathbf{k}}^{\mathrm{PC}}\rangle$ at the same energy $E_n= E_m$.}

\begin{figure}[htbp]
\centering
\includegraphics[scale=0.4]{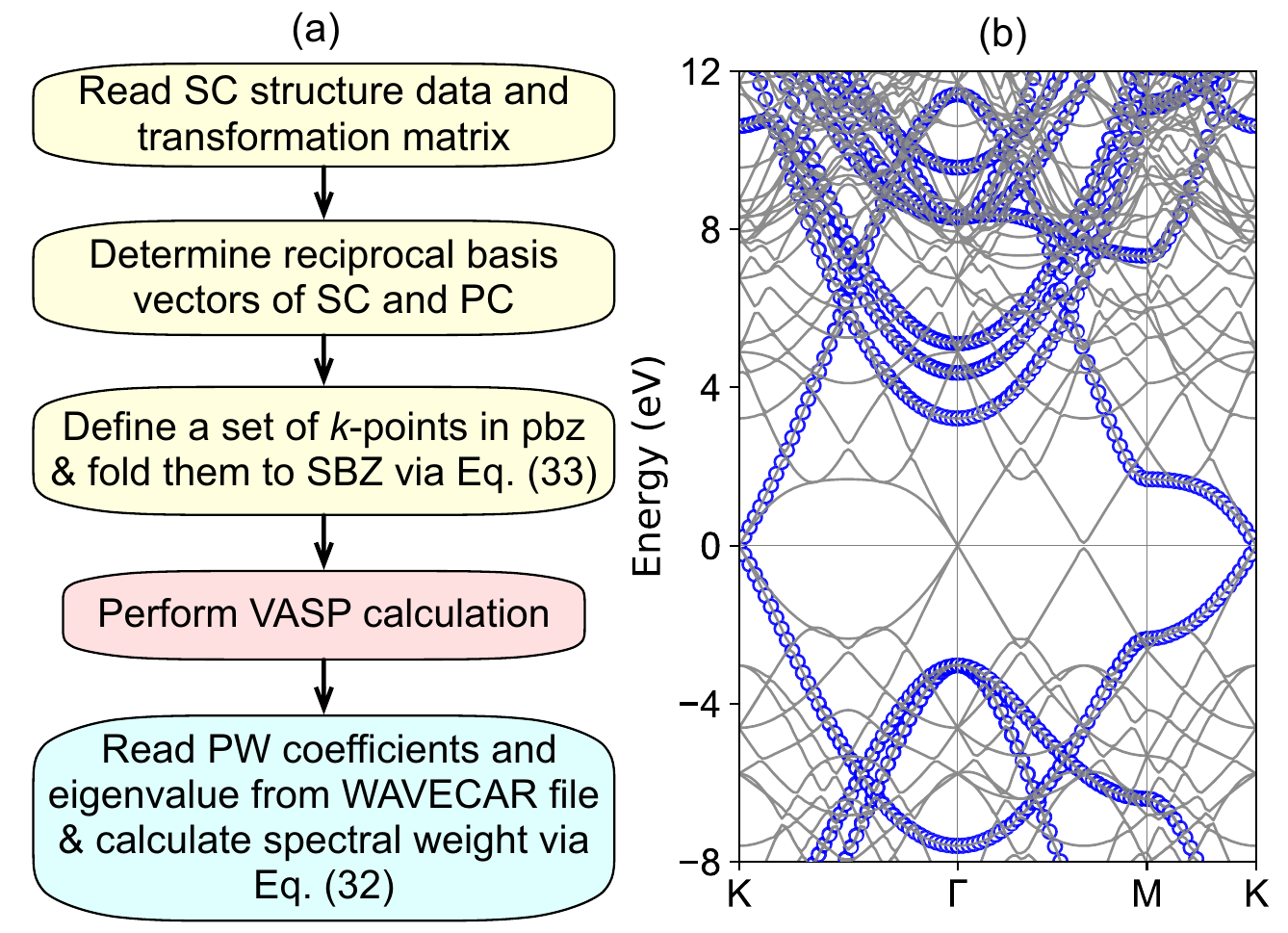}
\caption{\label{gebs}(Color online) (a) Workflow of the algorithm used in the band unfolding utility. (b) Band structure of 3$\times$3 graphene SC along the high-symmetry directions in pbz. The blue lines and red makers represent the band structure before and after applying the unfolding technique. The Fermi energy is set to zero.}
\end{figure}

\textcolor{black}{The workflow of band unfolding utility is schematically shown in Fig. \ref{gebs}(a). Three input files including the information of SC structure, the transformation matrix $\mathbf{M}$, and the selected $\mathbf{k}_{i}$ vectors in pbz are required to provide respectively. To compare the unfolded band structure of SC with the band structure of PC directly, the $\mathbf{k}_{i}$ vectors are generally sampled along the high-symmetry directions in pbz and then translated in the SC reciprocal space by the transformation as described in Eq. (\ref{k2K})}
\begin{equation}\label{k2K}
\mathbf{K}=\mathbf{M} \cdot \mathbf{k}_{i},
\end{equation}
\textcolor{black}{ where $\mathbf{K}$ and $\mathbf{k}_{i}$ are the scaled coordinates with respect to the SC and PC reciprocal basis vectors, respectively. After reading PW coefficients and eigenvalue of each state from the WAVECAR obtained by performing VASP calculation, the intricate supercell states can be unfolded back into the larger pbz by applying the unfolding technique via Eq. (\ref{ebs}). Finally the unfolded band can be visualized with the maker size proportional to the spectral weight $P_{\mathbf{K} m}$. 
From Fig. \ref{gebs}(b) it is clear that the folding of the bands into the smaller SBZ gives rise to quite sophisticated band structure. In contrast, one can gain more straightforward analysis once the supercell states are unfolded into the pbz despite the equivalence between the PC and the SC descriptions of a perfectly periodic material.}

\begin{figure}[htbp]
\centering
\includegraphics[scale=0.4]{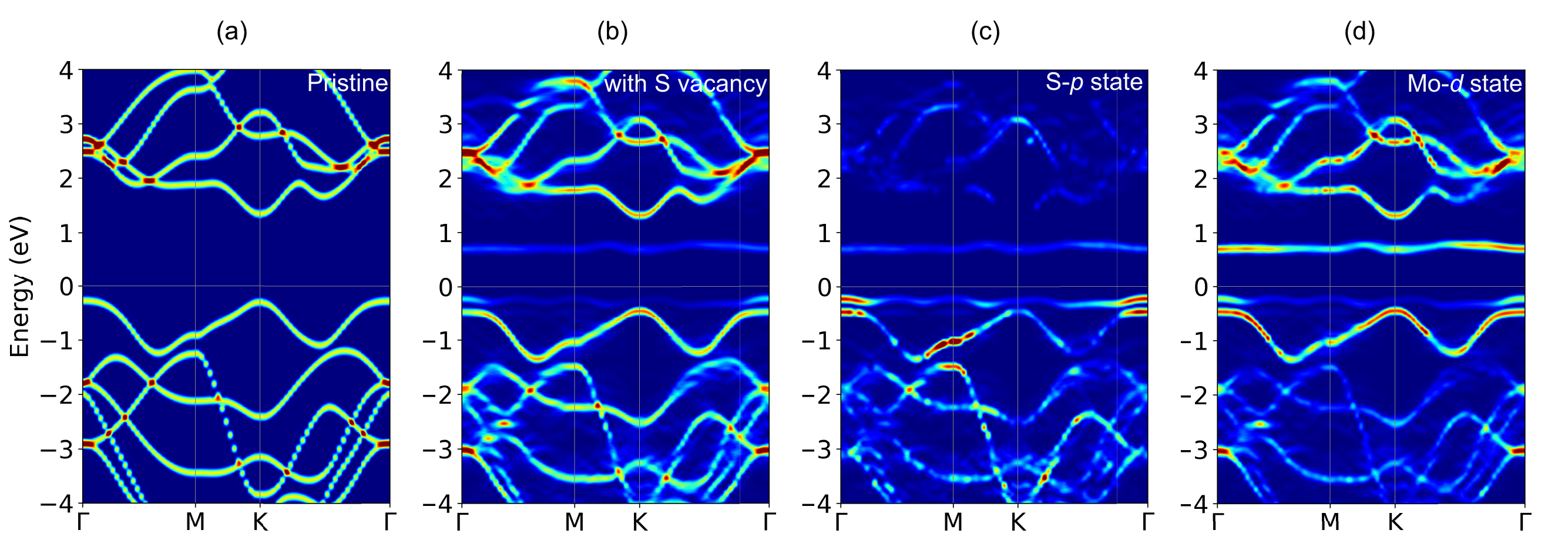}
\caption{\label{pebs}(Color online) Effective band structure of 4$\times$3 MoS$_2$ SC unfolded into the PC Brillouin zone through Eq. \ref{ebs} (a) without and (b) with a S vacancy.  Orbital-resolved effective band structure of (c) S-$p$ and (d) Mo-$d$ states in the defective SC. The  Fermi energy is set to zero.}
\end{figure}

\textcolor{black}{It is well known that intrinsic defects ﻿(vacancies, self-interstitials, and antisities) and unintentional impurities ﻿have important effects on the properties of semiconductors. As a typical case, we take the 4$\times$3 MoS$_2$ monolayer SC with one neutral sulfur vacancy as an example to demonstrate the role of intrinsic defect on the electronic structure of the pristine host. The calculated effective band structures of pristine and defective MoS$_2$ supercells in Figs. \ref{pebs} (a) and (b) respectively. By comparing these two, one can clearly find two nearly degenerated defect states existing in the fundamental band gap of MoS$_2$. Furthermore, the Bloch character close to the valence band edge is perturbed due to the presence of the sulfur vacancy. The orbital-resolved unfold band structures as show in Figs. (c) and (d) further demonstrate that these two defect states are mainly derived from Mo-$d$  and S-$p$ states respectively. }

\subsection{Linear Optical Properties}

The linear optical properties of semicondutors can be obtained from the frequency-dependent complex dielectric function $\varepsilon(\omega)$
\begin{equation}\label{epsilon_3D}
\varepsilon(\omega)=\varepsilon_{1}(\omega)+i \varepsilon_{2}(\omega),
\end{equation}
where $\varepsilon_{1}(\omega)$ and $\varepsilon_{2}(\omega)$ are the real and imaginary parts of the dielectric function, and $\omega$ is the photon frequency. Within the one-electron picture, the imaginary part of the dielectric function $\varepsilon_{2}(\omega)$ is obtained from the following equation \cite{Gajdos2006}:

\begin{equation}
\begin{aligned} \varepsilon_{2}(\omega)=& \frac{4 \pi^{2} e^{2}}{\Omega} \lim _{q \rightarrow 0} \frac{1}{q^{2}} \\ & \times \sum_{c, v, \mathbf{k}} 2 w_{\mathbf{k}} \delta\left(E_{c}-E_{v}-\omega\right)|\langle c|\mathbf{e} \cdot \mathbf{q}| v\rangle|^{2}, \end{aligned}
\end{equation}
where $\langle c|\mathbf{e} \cdot \mathbf{q}| v\rangle$ is the integrated optical transitions from the valence states ($v$) to the conduction states ($c$), $\mathbf{e}$ is the polarization direction of the photon and $\mathbf{q}$ is the electron momentum operator. The integration over $\mathbf{k}$ is performed by summation over special $k$-points with a corresponding weighting factor $w_{k}$. 
The real part of the dielectric function $\varepsilon_{1}(\omega)$ is obtained from the imaginary part $\varepsilon_{2}(\omega)$ based on the usual Kramers-Kronig transformation

\begin{equation}
\varepsilon_{1}(\omega)=1+\frac{2}{\pi} P \int_{0}^{\infty} \frac{\varepsilon_{\alpha \beta}^{(2)}\left(\omega^{\prime}\right) \omega^{\prime}}{\omega^{2}-\omega^{2}+i \eta} d \omega^{\prime},
\end{equation}
where $ P$ denotes the principle value and $\eta$ is the complex shift parameter. 
The frequency-dependent linear optical spectra, e.g., refractive index $n(\omega)$, extinction coefficient $\kappa(\omega)$, absorption coefficient $\alpha(\omega)$, energy-loss function $L(\omega)$, and reflectivity $R(\omega)$ can be calculated from the real $\varepsilon_{1}$($\omega$) and the imaginary $\varepsilon_{2}$($\omega$) parts \cite{Fox2002}:

\begin{equation}\label{o1}
n(\omega)=\left[\frac{\sqrt{\varepsilon_{1}^{2}+\varepsilon_{2}^{2}}+\varepsilon_{1}}{2}\right]^{\frac{1}{2}},
\end{equation}

\begin{equation}\label{o2}
k(\omega)=\left[\frac{\sqrt{\varepsilon_{1}^{2}+\varepsilon_{2}^{2}}-\varepsilon_{1}}{2}\right]^{\frac{1}{2}},
\end{equation}

\begin{equation}\label{o3}
\alpha(\omega)= \frac{\sqrt{2} \omega}{c}\left[\sqrt{\varepsilon_{1}^{2}+\varepsilon_{2}^{2}}-\varepsilon_{1}\right]^{\frac{1}{2}},
\end{equation}

\begin{equation}\label{o4}
L(\omega)=\operatorname{Im}\left(\frac{-1}{\varepsilon(\omega)}\right)=\frac{\varepsilon_{2}}{\varepsilon_{1}^{2}+\varepsilon_{2}^{2}},
\end{equation}

\begin{equation}\label{o5}
R(\omega)=\frac{(n-1)^{2}+k^{2}}{(n+1)^{2}+k^{2}}.
\end{equation}

\begin{figure}[htbp]
\centering
\includegraphics[scale=0.37]{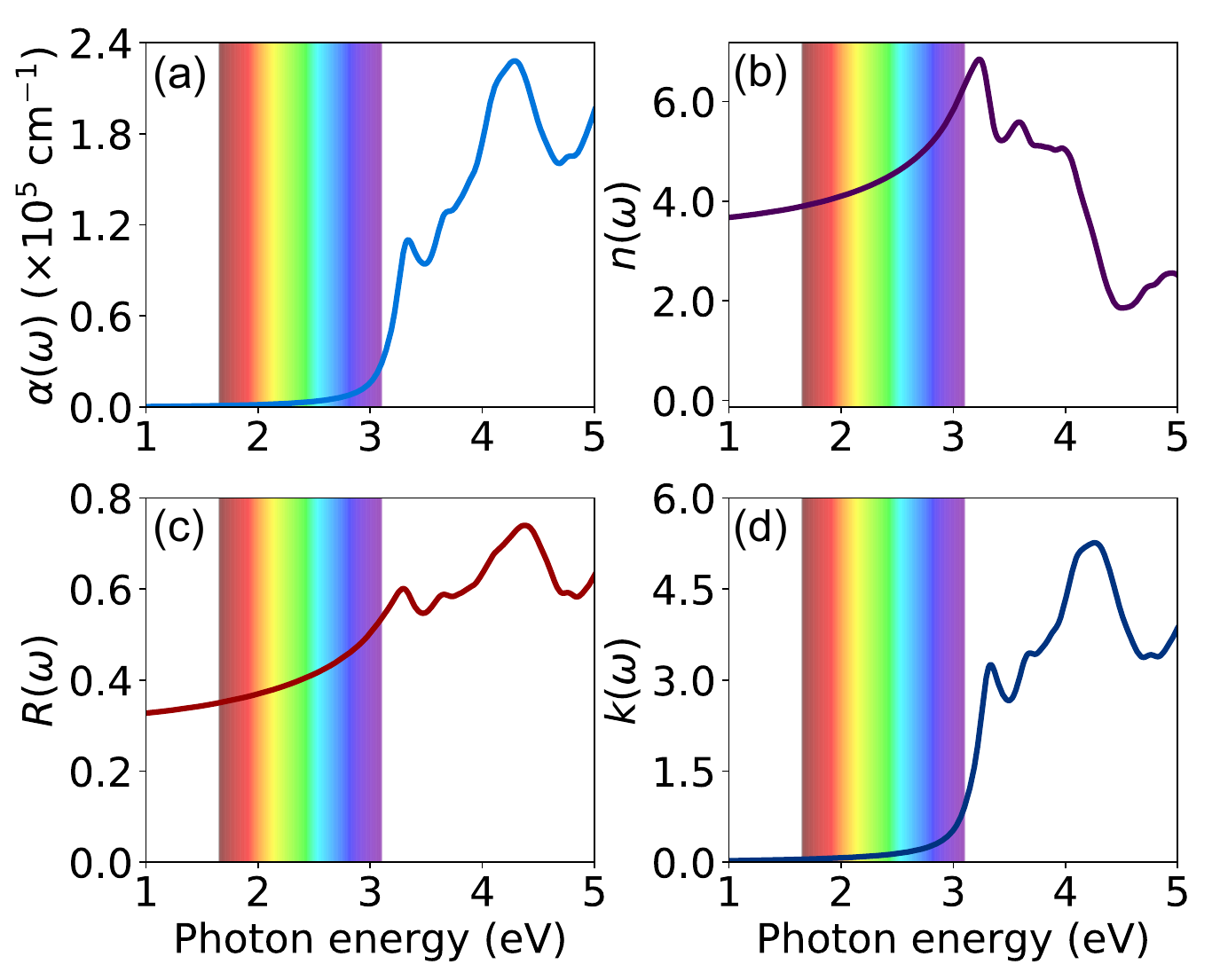}
\caption{\label{Optical}(Color online) G$_0$W$_0$-BSE calculated (a) absorption coefficient, (b) refractive index, (c) reflectivity and (d) extinction coefficient of silicon. \textcolor{black}{The visible light region is highlighted by vertical color lines.}}
\end{figure}

In Figure \ref{Optical} we present the linear optical spectra of silicon as determined by solving the Bethe-Salpeter Equation (BSE) on the top of G$_0$W$_0$ approximation.
One can find that the absorption coefficient become significant only after 3.0 eV. This is because silicon has an indirect band gap, resulting in a low absorption coefficient in the visible region. \textcolor{black}{Since the GW approximation includes the exchange and correlation effects in a self-energy term dependent on the one particle Green's function G and the dynamically screened Coulomb interaction W, it can correct the one electron eigenvalues obtained from DFT within a many-body quasiparticle framework \cite{Hedin1965,Fuchs2007}. Furthermore, the errors originated from the lack of ladder diagrams in determining W can be included through solution of the Bethe-Salpeter equation (BSE) \cite{Onida2002}. It \textcolor{black}{could} be expected that the GW-BSE} calculated optical properties yield better agreement with experiment. In the single-shot G$_0$W$_0$ approximation, the one-electron Green's function G is self-consistently updated within a single iteration, while the screened Coulomb interaction W is fixed at its initial value.  

\begin{figure}[htbp]
\centering
\includegraphics[scale=0.37]{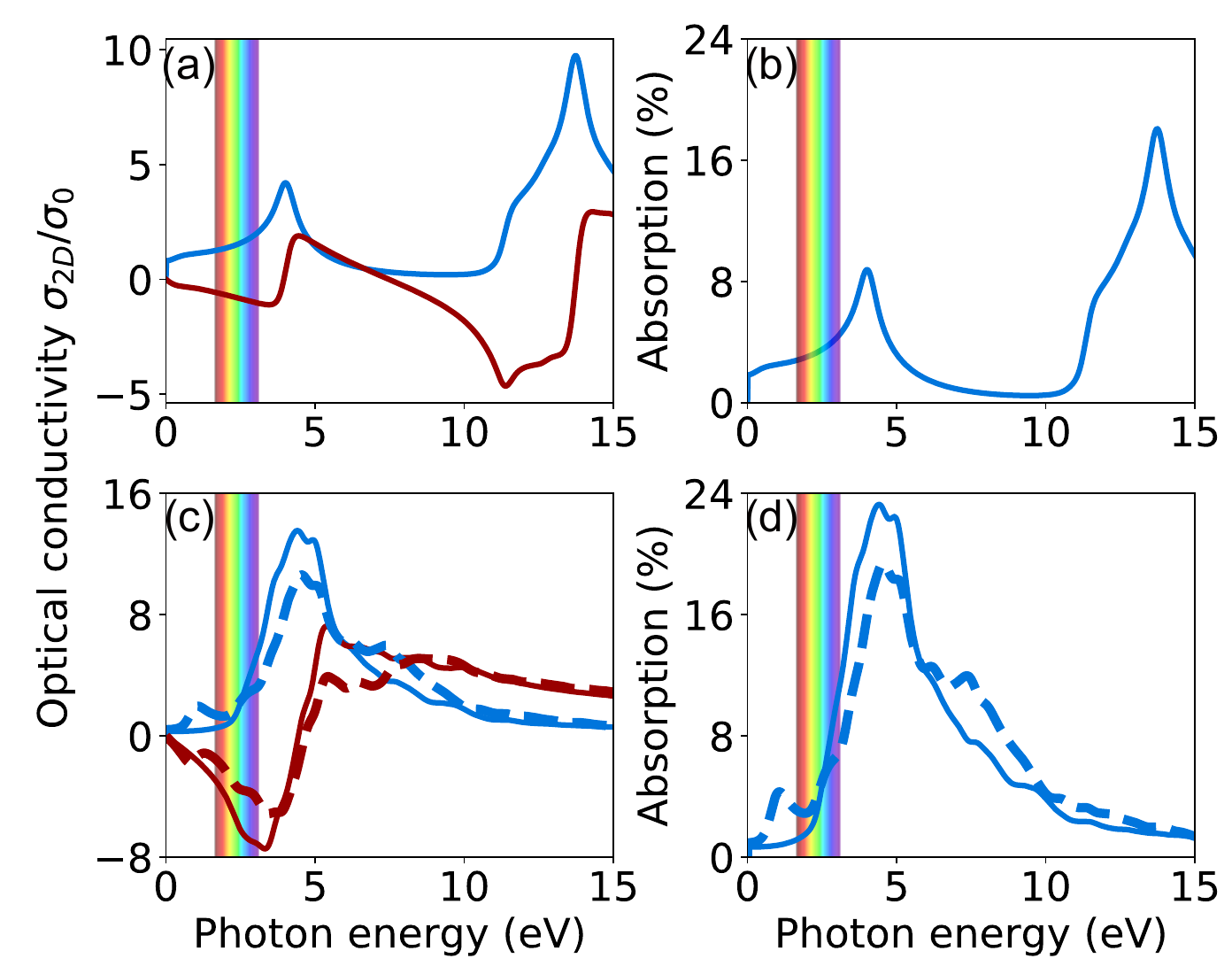}
\caption{\label{Optical_2D}(Color online) Real (blue line) and imaginary (red line) parts of frequency-dependent optical conductivity $\sigma_{2D}(\omega)$ for (a) graphene and (c) phosphorene  [in units of $\sigma_0=e^{2} /(4 \hbar)$]. Absorption spectra $A(\omega)$ of (b) graphene and (d) phosphorene. The incident light polarized along the armchair and zigzag directions of phosphorene are presented by solid and dashed lines respectively. \textcolor{black}{The visible light region is highlighted by vertical color lines.}}
\end{figure}

It should be pointed out that the Eqs. (\ref{o1})-(\ref{o5}) are not well-defined for low-dimensional materials since the dielectric function is not straightforward and depends on the thickness of the vacuum layer when the low-dimensional systems are simulated using a periodic stack of layers with sufficiently large interlayer distance $L$ to avoid artificial interactions between the periodic images of the 2D sheet crystals in the standard DFT calculations \cite{Hueser2013,Cudazzo2011}. To avoid the thickness problem, the optical conductivity $\sigma_{2D}(\omega)$ is used to characterize the optical properties of 2D sheets. Based on the Maxwell equation, the 
3D optical conductivity can be expressed as \cite{Matthes2014}

\begin{equation}
\sigma_{3D}(\omega)=i[1-\varepsilon(\omega)] \varepsilon_{0} \omega,
\end{equation} 
where $\varepsilon(\omega)$ is frequency-dependent complex dielectric function given in (\ref{epsilon_3D}), $\varepsilon_{0}$ is the permittivity of vacuum and $\omega$ is the frequency of incident wave. The in-plane 2D optical conductivity is directly related to the corresponding $\sigma_{3D}(\omega)$ component through the equation \cite{Matthes2014,Matthes2016}

\begin{equation}\label{eq_conduc_2D}
\sigma_{2 D}(\omega)=L \sigma_{3 D}(\omega),
\end{equation}
where $L$ is the slab thickness in the simulation cell. The normalized reflectance $R(\omega)$, transmittance $T(\omega)$ and absorbance $A(\omega)$ are independent of the light polarization for a freestanding 2D crystal sheet when normal incidence is assumed \cite{Matthes2014,Matthes2016},

\begin{equation}\label{eq_op_2D}
\begin{aligned} R &=\left|\frac{\tilde{\sigma} / 2}{1+\tilde{\sigma} / 2}\right|^{2}, \\ T &=\frac{1}{|1+\tilde{\sigma} / 2|^{2}}, \\ A &=\frac{\operatorname{Re} \tilde{\sigma}}{|1+\tilde{\sigma} / 2|^{2}},  \end{aligned}
\end{equation}
where $\tilde{\sigma}(\omega)=\sigma_{2 \mathrm{D}}(\omega) / \varepsilon_{0} c$ is the normalized conductivity ($c$ is the speed of light).  Since the interband contribution is only considered, the formula (\ref{eq_op_2D}) is valid for semiconducting and insulating 2D crystals with restriction of $A+T+R=1$. Generally, the reflectance of 2D sheets is extremely small and the absorbance can be approximated by the real part of $\tilde{\sigma}(\omega)$, namely, $A(\omega)=\operatorname{Re} \sigma_{2 D}(\omega)/{\varepsilon_{0} c}$. To demonstrate this functionality, the PBE-calculated linear optical spectra of freestanding graphene and phosphorene monolayers are displayed in Figure \ref{Optical_2D}. Our results are in good agreement with the available theoretical optical curves \cite{Matthes2014,Matthes2016,Torbatian2018}.

\subsection{Joint Density of States}
\begin{figure}[htbp]
\centering
\includegraphics[scale=1.25]{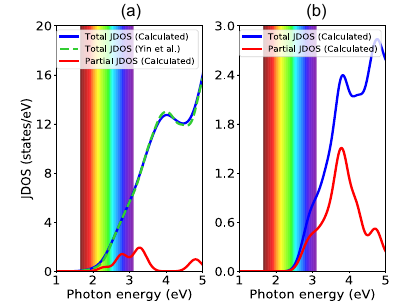}
\caption{\label{JDOS}(Color online) Calculated joint  density of states for (a) CH$_3$NH$_3$PbI$_3$ and (b) Si. Blue and purple lines represent the total and partial joint density of states respectively. \textcolor{black}{The visible light region is highlighted by vertical color lines.}}
\end{figure}

For a semiconductor, the optical absorption in direct band-to-band transitions is proportional to \cite{dresselhaus2001}

\begin{equation}
\frac{2 \pi}{\hbar} \int_{\mathrm{BZ}}\left|\left\langle v\left|\mathcal{H}^{\prime}\right| c\right\rangle\right|^{2} \frac{2}{(2 \pi)^{3}} \delta\left(E_{c}(\mathbf{k})-E_{v}(\mathbf{k})-\hbar \omega\right) d^{3} k,
\end{equation}
where $\mathcal{H}^{\prime}$ is the perturbation associated with the light wave and $\left\langle v\left|\mathcal{H}^{\prime}\right| c\right\rangle$ is the transition matrix from states in the valence band (VB) to states in the conduction band (CB); $\delta$ is the Dirac delta function which switches on this contribution when a transition occurs from one state to another, i.e., $E_{c}(\mathbf{k})-E_{v}(\mathbf{k})=\hbar \omega$. The factor 2 stems from the spin degeneracy. The integration is over the entire BZ. The matrix elements vary little within the BZ. Therefore, we can pull these out in front of the integral and obtain

\begin{equation}
\frac{2 \pi}{\Omega \hbar}\left|\left\langle v\left|\mathcal{H}^{\prime}\right| c\right\rangle\right|^{2} \cdot \int \frac{2 \Omega}{(2 \pi)^{3}} \delta\left(E_{c}(\mathbf{k})-E_{v}(\mathbf{k})-\hbar \omega\right) \mathrm{d}^{3} k,
\end{equation}
where $\Omega$ is the volume of the lattice cell,  and the factor $\Omega/{(2 \pi)^{3}}$ normalizes the $\mathbf{k}$ vector density within the Brillouin zone. The second term is the joint density of states (JDOS). After sum over all states within the first Brillouin zone and all possible transitions initiated by photons with a certain energy $\hbar \omega$ between valence and conduction bands, we obtain

\begin{equation}\label{ejdos}
\begin{aligned} j(\omega)=\sum_{v,c} \frac{\Omega}{4 \pi^{3}} \int \delta\left(E_{c}(\mathbf{k})-E_{v}(\mathbf{k})-\hbar \omega\right)  d^{3} k\\ = 2\sum_{v,c,\mathbf{k}} w_{\mathbf{k}} \delta\left(E_{c}(\mathbf{k})-E_{v}(\mathbf{k})-\hbar \omega\right), \end{aligned}
\end{equation}
where $c$ and $v$ belong respectively to the valence and conduction bands, $E(\mathbf{k})$ are the eigenvalues of the Hamiltonian, and $w_{\mathbf{k}}$ are weighting factors. The Dirac Delta function in Eq. (\ref{ejdos}) can be numerically approximated by means of a normalized Gaussian function:

\begin{equation}
G(\omega)=\frac{1}{\sigma \sqrt{2\pi}} e^{-\left(E_{\mathbf{k}, n^{\prime}}-E_{\mathbf{k}, n}-\hbar \omega\right)^{2} / 2\sigma^{2}},
\end{equation}
where $\sigma$ is the broadening parameter. To demonstrate this functionality, we show the calculated total and partial JDOS for CH$_3$NH$_3$PbI$_3$ and Si in Fig. \ref{JDOS}. Clearly, the calculated JDOS for CH$_3$NH$_3$PbI$_3$ is in excellent agreement with previous data \cite{Yin2014}. It should be pointed that the total JDOS include all possible interband transitions from all the valence to all the conduction bands according to Eq. (\ref{ejdos}); while the partial JDOS consider only the interband transitions from the highest VB to the lowest CB. 

\subsection{Transition Dipole Moment}
\begin{figure}[htbp]
\centering
\includegraphics[scale=0.48]{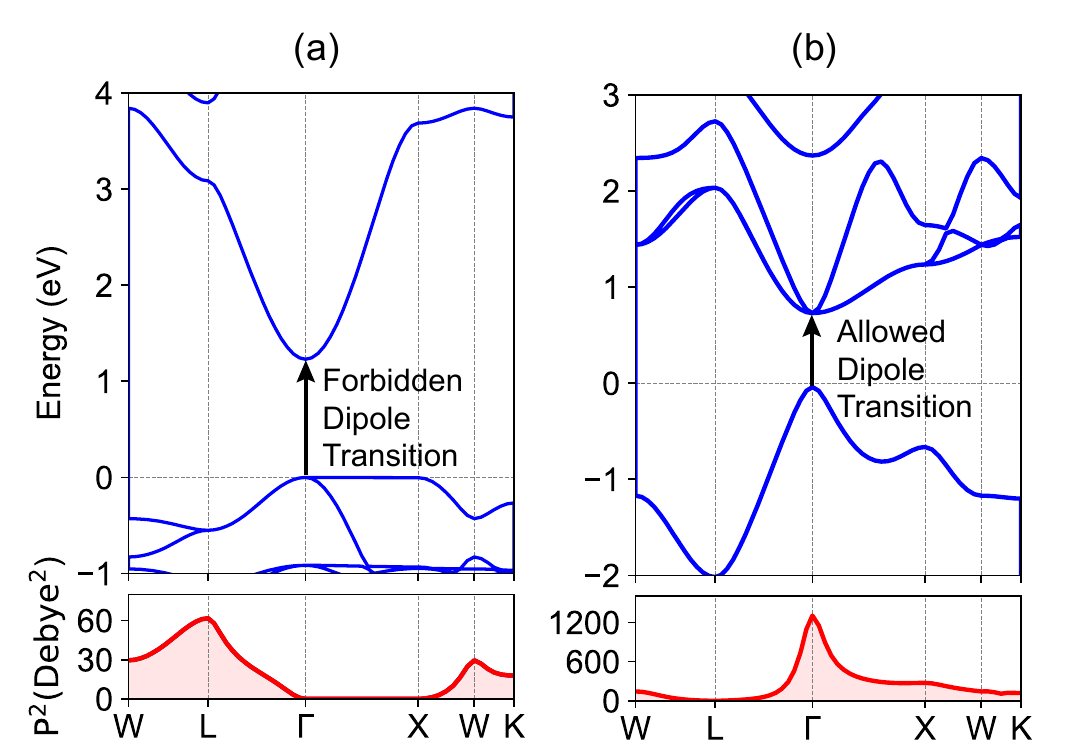}
\caption{\label{TDM}(Color online) Calculated band structure (top panel) and transition dipole moment (bottom panel) for (a) Cs$_2$AgInCl$_6$ and (b) Cs$_2$InBiCl$_6$.}
\end{figure}

The transition dipole moment (TDM) or dipole transition matrix elements $\mathrm{P}_{a \rightarrow b}$, is the electric dipole moment associated with a transition between the initial state $a$ and the final state $b$ \cite{wiki}:

\begin{equation}
\mathrm{P}_{a \rightarrow b}=\left\langle\psi_{b}|\mathbf{r}| \psi_{a}\right\rangle=\frac{i \hbar}{\left(E_{b}-E_{a}\right) m}\left\langle\psi_{b}|\mathbf{p}| \psi_{a}\right\rangle,
\end{equation}
where $\psi_{a}$ and $\psi_{b}$ are energy eigenstates with energy $E_{a}$ and $E_{b}$; $m$ is the mass of the electron. In general the TDM is a complex vector  that includes the phase factors associated with the two states. Its direction gives the polarization of the transition, which determines how the system will interact with an electromagnetic wave of a given polarization, while the sum of the squares of TDM, $\emph{P}^2$, give the transition probabilities between \textcolor{black}{the two} states.  In Fig. \ref{TDM} we provide some specific examples to illustrate its use. It is seen that the calculated TDM amplitude is zero for transition between the CBM and VBM at the $\Gamma$ point in Cs$_2$AgInCl$_6$, implying no optical absorption between these two states. On the other hand, the excellent optical absorption between CBM and VBM is predicted in Cs$_2$InBiCl$_6$ when Ag atom is substituted by Bi. These findings are in good agreement with previous theoretical results \cite{Meng2017}.

\subsection{$d$-Band Center} 
The $d$-band center model of Hammer and N{\o}rskov is widely used in understanding and predicting catalytic activity on transition metal surfaces. The \textcolor{black}{main idea} underlying the theory is that the binding energy of an adsorbate to a metal surface is largely dependent on the electronic structure of the surface itself. In this model, the band of $d$-states participating in the interaction is approximated by the center of the $d$-band $\varepsilon_{d}$ \cite{Norskov2014}:

\begin{equation}\label{dbc}
\varepsilon_{\mathrm{d}}=\frac{\int_{-\infty}^{\infty} n_{\mathrm{d}}(\varepsilon) \varepsilon d \varepsilon}{\int_{-\infty}^{\infty} n_{\mathrm{d}}(\varepsilon) d \varepsilon},
\end{equation}
where $n_{\mathrm{d}}$ and $\varepsilon$ are projected-DOS and energy of transition metal $d$ states.
According to this model, the adsorption energy on transition metal surface correlates the upward shift of $d$-band center with respect to the Fermi energy. A stronger upward shift indicates the possibility of the formation of a larger number of empty anti-bonding states, leading to a stronger binding energy \cite{Hammer1995,Hammer2000,Norskov2014}. It may be worth mentioning here that the position of $d$-band center linearly upshifts with increasing the number of empty states above the Fermi level. Therefore, one can specify the integral upper limit in Eq. (\ref{dbc}) to calculate $d$-band center by using VASPKIT.

\subsection{Thermo Energy Correction}
Gibbs free energy plays a crucial role in catalysis reaction. The equations used for calculating thermochemical data for gases in VASPKIT is equivalent to those in Gaussian program \cite{McQuarrie1999,Ochterski2000}. The Gibbs free energy $G$ is given by 
\begin{equation}\label{gibbs}
G=H-T S,
\end{equation}
where $H$, $T$ and $S$ represent enthalpy, temperature and entropy respectively. \textcolor{black}{The enthalpy $H$ in Eq. (\ref{gibbs}) can be written as $H=U+PV$}. Both internal thermal energy $U$ and entropy $S$ have included the contributions from translational, electronic, rotational and vibrational motions as well as zero-point energy (ZPE) of molecules. Moreover, to calculate correctly when the number of moles (labeled $N$) of a gas changes during the course of a reaction, the Gibbs free energy has also included $\Delta P V=\Delta N R T$, where $R$ is molar gas constant. It is worth mentioning that only the modes with real vibrational frequencies are considered and the model with imaginary one are ignored during the calculations of the vibration contributions. Specifically, for linear (non-linear) molecules containing $n$ atoms, the degree of vibrational freedom is 3$n$ - 5 (3$n$ - 6). VASPKIT \textcolor{black}{neglects} the smallest 5 (6) frequencies. We take oxygen molecular as an example to calculate its free energy  at 298.15 K using the corrected algorithm mentioned above. It is found that the calculated correction to free energy of O$_2$ molecule is -0.4467 eV, which is very close to the experimental data of -0.4468 eV at 298.15 K and normal atmospheric pressure \cite{Chase1998}. And the thermo correction result from VASPKIT is exactly the same with that from Gaussian program by setting the same molecular structure and frequencies.

Unlike gas molecules, when the adsorbed molecules form chemical bonds with substrate, their translational and rotational freedom will be constrained. Consequently, the contributions from translation and rotation to entropy and enthalpy are significantly reduced turn into vibrational modes (at least at low temperatures-at higher temperatures, they might become frustrated translational or frustrated rotational). One common method is to attribute the translational or rotational part of the contribution to vibration, that is, the 3$n$ vibrations of the surface-adsorbing molecules (except the imaginary frequency) are all used to calculate the correction of the thermo energy \cite{Norskov2014}.
Considering that a small vibration mode makes a large contribution to entropy. It is very likely that a small vibration frequency will lead to abnormal entropy and free energy correction. Thus, VASPKIT allows to specify a threshold value which defines the lower limit of frequencies. For example, if a threshold value of 50 cm$^{-1}$ is adopted, implying that the frequencies below 50 cm$^{-1}$ are approximately equal to 50 cm$^{-1}$ during the calculations of the vibration contributions to the adsorbed molecular free energy correction.

\subsection{Molecular Dynamics}

The molecular dynamics (MD) \textcolor{black}{describes}  how the atoms in a material move as a function of time, \textcolor{black}{and helps us to understand} the structural, dynamical and thermodynamical properties of complex systems. It has been successfully applied to gases, liquids, and ordered and disordered solids. In addition to equation of state, mean square displacement (MSD), velocity auto-correlation function (VACF), \textcolor{black}{phonon vibrational density of states (VDOS)} and pair correlation function (PCF) are the most important quantities enabling us to determine various properties of interest in MD simulations.

The MSD is a measure of the deviation of the position of a particle with respect to a reference position over time. It can help to determine whether the ion is freely diffusing, transported, or bound.  It is defined as

\begin{equation}
MSD(m)=\frac{1}{N_{\text {particles}}} \sum_{i=1}^{N_{\text {paricles}}} \frac{1}{N-m} \sum_{k=0}^{N-m-1}\left(\mathbf{r}_{i}(k+m)-\mathbf{r}_{i}(k)\right)^{2},
\end{equation}
where $\mathbf{r}_{i}(t)$ is the position of atom $i$ after $t$ time of simulation. $N_{\text {particles}}$ and $N$ are the total number of atoms and total frames respectively.  According to this definition, the MSD is averaged over all windows of length $m$ and over all selected particles. An alternative method which can efficiently calculate MSD \textcolor{black}{was}  proposed based on the Fast Fourier Transform (FFT) algorithm in Refs. \cite{Kneller1995,Rog2003} and references therein. If the system stays in the solid state, the MSD oscillates around a constant value. This means that all the atoms are confined to certain positions. For a liquid, however, atoms will move indefinitely and the MSD continues to increase linearly with time. This implies that sudden changes in the MSD with time are indicative of melting, solidification, phase transition, and so on. In addition, the calculation of MSD is the standard way to estimate the parameters of movement, such as the diffusion coefficients from MD simulations. 

The VACF is another way of checking the movement type of atoms. It is a value that basically tells until when the particle remembers its previous movements. Like  the MSD, it is a time-averaged value, defined over a delay domain. The normalized VACF is defined as 
\begin{equation}
c(t)=\frac{\sum_{i=1}^{N}\left\langle\mathbf{v}_{i}(t) \cdot \mathbf{v}_{i}(0)\right\rangle}{\sum_{i=1}^{N}\left(\mathbf{v}_{i}(0)\right)^{2}},
\end{equation}
where $\mathbf{v}_{i}(t)$ is the velocity of the \textcolor{black}{$i$-$th$} atom at time $t$. The bracket represents a time average over the history of the particle, i.e., all the values of t.
\textcolor{black}{The total velocity autocorrelation function $C(t)$ is deﬁned as the mass-weighted sum of the atom velocity autocorrelation functions \cite{Lin2003}}

\begin{equation}
C(t)=\sum_{j=1}^{N} m_{j} c_{j}(t),
\end{equation}
\textcolor{black}{where $c_{j}(t)$ is the velocity autocorrelation of atom $j$}. The optical and thermodynamical properties of materials depend on VDOS which can be obtained from the Fourier transform of the VACF under the harmonic approximation \cite{DICKEY1969,Lin2003},

\begin{equation}
f(\omega)=\mathcal{F}[\gamma(t)]=\frac{1}{k_{B} T}\int_{-\infty}^{\infty} \gamma(t) e^{-i{\omega}t} dt,
\end{equation}
where $\omega$ is  the vibrational frequency, $\mathcal{F}$ is the Fourier transform operator, $k_{B}$ is the Boltzmann constant and $T$ is the absolute temperature. 

The PCF g($r$) describes how atoms are distributed in a thin shell at a radius $r$ from an arbitrary atom in the material. It is useful not only for studying the details of the system but also to obtain accurate values for the macroscopic quantities such as the potential energy and pressure. This quantity can be obtained by summing the number of atoms found at a given distance in all directions from a particular atom:

\begin{equation}
g(r)=\frac{d N / N}{d V / V}=\frac{1}{4 \pi r^{2}} \frac{1}{N \rho} \sum_{i=1}^{N} \sum_{j \neq i}^{N}\left\langle\delta\left(r-\left|\mathbf{r}_{i}-\mathbf{r}_{j}\right|\right)\right\rangle,
\end{equation}
where $r$ is the radial distance. $\rho$ is the average density of the entire material. The normalization via the density ensures that for large distances the radial distribution approaches unity. The partial radial distribution between two elements is calculated as

\begin{equation}
g_{A B}(r)=\frac{1}{4 \pi r^{2}} \frac{N}{\rho N_{A} N_{B}} \sum_{i \in A} \sum_{j \in B, j \neq i}^{N}\left\langle\delta\left(r-\left|\mathbf{r}_{i}-\mathbf{r}_{j}\right|\right)\right\rangle
\end{equation}.

\begin{figure}[htbp]
\centering
\includegraphics[scale=0.38]{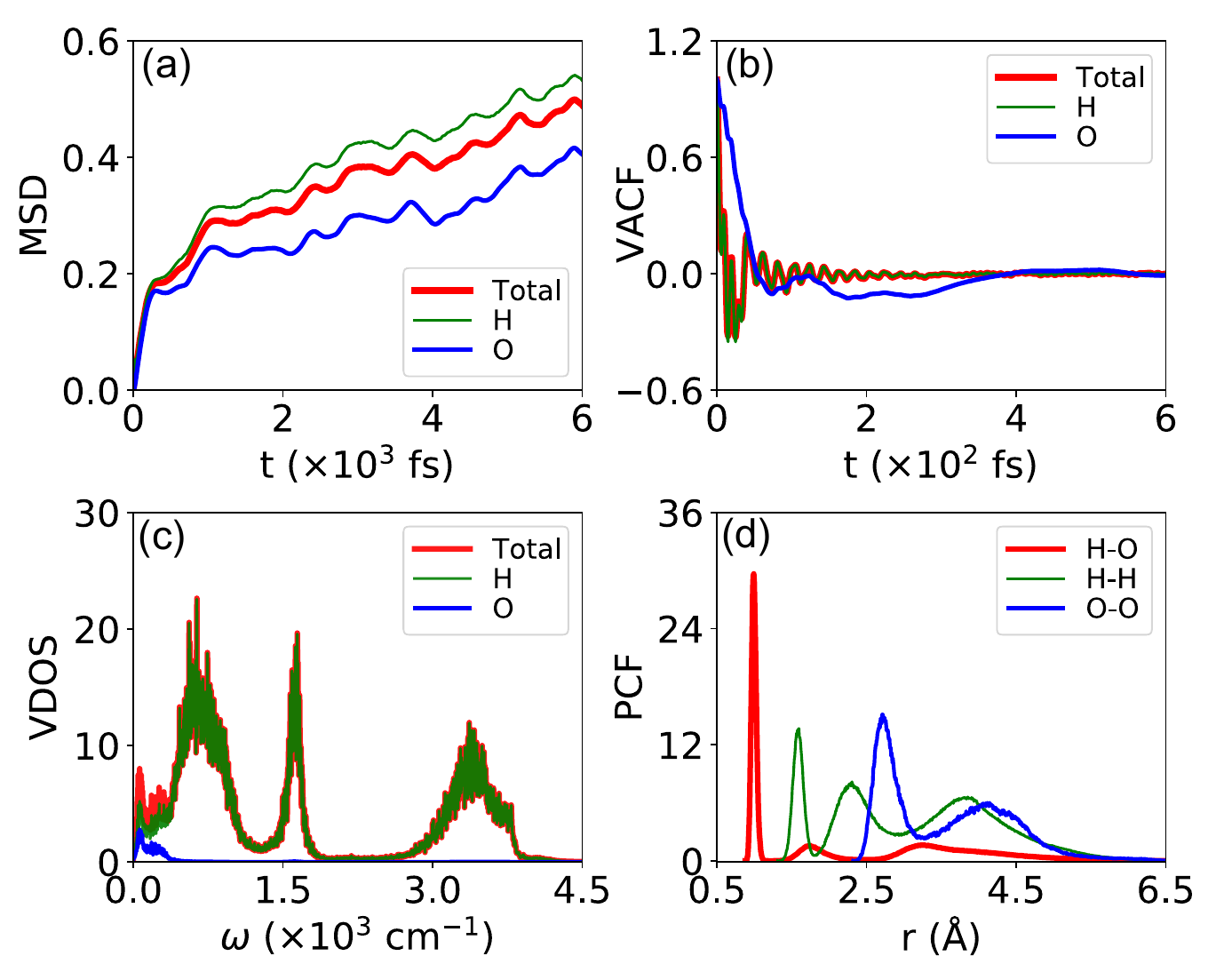}
\caption{\label{md}(Color online) Calculated (a) MSD, (b) VACF, (c) VDOS and (d) PCF of liquid water at 400 K obtained from MD simulations.}
\end{figure}

As an illustrated example, Figure \ref{md} shows the PBE calculated MSD, VACF, VDOS and PCF for liquid water at 400 K processed by the MD utility. Overall, our result is in good agreement with available experimental and theoretical results  \cite{Soper2008,Imoto2013}.

\section{\textcolor{black}{High-throughput capabilities}}

\textcolor{black}{VASPKIT also provides a light-weight high-throughput interface. As such it can advantageously be part of bash scripts, taking full advantage of bash capabilities (variables, loops, conditions, etc.) to batch performing  pre- and post-processing.  An easy-to-follow user manual is available at \href{https://vaspkit.com/tutorials.html}{https://vaspkit.com/tutorials.html}.
The syntax is designed as simple as possible. For instance, to generate KPOINTS files in a series of subfolders, the syntax is}

\begin{lstlisting}[language=bash]
RootPath=`pwd`
for dir in *
do 
   echo $dir
   cd $RootPath/$dir    
   vaspkit -task 102 -kpr 0.04   
done
\end{lstlisting}

\section{\textcolor{black}{Limitations and future capabilities}}

\textcolor{black}{Currently, VASPKIT only deals with the raw data calculated using the VASP code. This program will be extended to support other \emph{ab-initio} packages in the future version. In addition, the data visualization and plotting utility based on Python and Matplotlib will be also implemented.}

\section{Summary}

In summary, VASPKIT is a user friendly toolkit that can be easily \textcolor{black}{employed} to perform initial setup for calculations and post-processing analysis to derive \textcolor{black}{a good many} material properties from the raw data generated by VASP code. We have demonstrated its capability through illustrative examples.  VASPKIT provides command-line interface for the purpose of performing high-throughput calculations. It remains under development, and further functionality, including closer support for other codes, \textcolor{black}{is readily to be implemented. With new features being added}, we hope that VASPKIT will become an even more attractive toolkit contributing to efficient development and utilization of electronic structure theory.

\section{Declaration of competing interest}
The authors declare that they have no known competing financial interests or personal relationships that could have appeared to influence the work reported in this paper.

\section{Acknowledgments}

We acknowledge other contributors (in no particular order) including Peng-Fei Liu, Xue-Fei Liu, Zhao-Fu Zhang, Tian Wang, Dao-Xiong Wu, Ya-Chao Liu, Jiang-Shan Zhao and Qiang Li. We gratefully acknowledge helpful discussions with Zhe-Yong Fan, Qi-Jing Zheng and Ming-Qing Liao. We also thank various researchers around the world for reporting bugs and suggesting features, which have lead to significant improvements in the accuracy and robustness of the package. V.W. gratefully appreciates Yoshiyuki Kawazoe and Shigenobu Ogata for their invaluable support. V.W. also thanks The Youth Innovation Team of Shaanxi Universities. 

\appendix
\section{Elastic stiffness tensor matrix and strain modes  for bulk crystal systems}

\begin{description}
\item [{1. Triclinic System (Space group numbers: 1-2)}]~
\end{description}

There are 21 independent elastic constants. $C_{11}$, $C_{12}$, $C_{13}$, $C_{14}$, $C_{15}$, $C_{16}$, $C_{22}$, $C_{23}$, $C_{24}$, $C_{25}$, $C_{26}$, $C_{33}$, $C_{34}$, $C_{35}$, $C_{36}$, $C_{44}$, $C_{45}$, $C_{46}$, $C_{55}$, $C_{56}$ and $C_{66}$  

The elastic stiffness tensor matrix is expressed by
\begin{equation}
C_{ij}=\left(\begin{array}{llllll}
C_{11} & C_{12} & C_{13} & C_{14} & C_{15} & C_{16} \\
C_{12} & C_{22} & C_{23} & C_{24} & C_{25} & C_{26} \\
C_{13} & C_{23} & C_{33} & C_{34} & C_{35} & C_{36} \\
C_{14} & C_{24} & C_{34} & C_{44} & C_{45} & C_{46} \\
C_{15} & C_{25} & C_{35} & C_{45} & C_{55} & C_{56} \\
C_{16} & C_{26} & C_{36} & C_{46} & C_{56} & C_{66}
\end{array}\right).
\end{equation}

\begin{table*}[htbp]
\centering
\caption{List of strain modes and the derived elastic constants for \textbf{triclinic} system used in VASPKIT based on energy-strain approach.}
\begin{tabular}{ccc}
\hline
Strain index & Strain vector $\boldsymbol{\varepsilon}$ & Elastic energy $\frac{\Delta E}{V}$  \\
\hline
1   &    $(\delta, 0, 0, 0, 0, 0)$       &      $\frac{1}{2} C_{11} \delta^{2}$ \\
2   &    $(0, \delta, 0, 0, 0, 0)$       &      $\frac{1}{2} C_{22} \delta^{2}$ \\
3   &    $(0, 0, \delta, 0, 0, 0)$       &      $\frac{1}{2} C_{33} \delta^{2}$ \\
4   &    $( 0, 0, 0,\delta, 0, 0)$       &      $\frac{1}{2} C_{44} \delta^{2}$ \\
5   &    $(0, 0, 0, 0,\delta,  0)$       &      $\frac{1}{2} C_{55} \delta^{2}$ \\
6   &    $(0, 0, 0, 0, 0,\delta )$       &      $\frac{1}{2} C_{66} \delta^{2}$ \\
7   &    $(\delta, \delta,0,0,0,0)$  & $\left(\frac{C_{11}}{2}+C_{12}+\frac{C_{22}}{2}\right) \delta^{2}$ \\
8   &    $(\delta,0, \delta,0,0,0)$  & $\left(\frac{C_{11}}{2}+C_{13}+\frac{C_{33}}{2}\right) \delta^{2}$ \\
9   &    $(\delta,0, 0,\delta,0,0)$  & $\left(\frac{C_{11}}{2}+C_{14}+\frac{C_{44}}{2}\right) \delta^{2}$ \\
10   &    $(\delta,0,0,0,\delta,0)$  & $\left(\frac{C_{11}}{2}+C_{15}+\frac{C_{55}}{2}\right) \delta^{2}$ \\
11   &    $(\delta,0,0,0,0,\delta)$  & $\left(\frac{C_{11}}{2}+C_{16}+\frac{C_{66}}{2}\right) \delta^{2}$ \\
12   &    $(0,\delta, \delta,0,0,0)$  & $\left(\frac{C_{22}}{2}+C_{23}+\frac{C_{33}}{2}\right) \delta^{2}$ \\
13   &    $(0,\delta,0,\delta,0,0)$  & $\left(\frac{C_{22}}{2}+C_{24}+\frac{C_{44}}{2}\right) \delta^{2}$ \\
14   &    $(0,\delta,0,0,\delta,0)$  & $\left(\frac{C_{22}}{2}+C_{25}+\frac{C_{55}}{2}\right) \delta^{2}$ \\
15   &    $(0,\delta,0,0,0,\delta)$  & $\left(\frac{C_{22}}{2}+C_{26}+\frac{C_{66}}{2}\right) \delta^{2}$ \\
16   &    $(0,0,\delta,\delta,0, 0)$  & $\left(\frac{C_{33}}{2}+C_{34}+\frac{C_{44}}{2}\right) \delta^{2}$ \\
17   &    $(0,0,\delta,0,\delta,0)$  & $\left(\frac{C_{33}}{2}+C_{35}+\frac{C_{55}}{2}\right) \delta^{2}$ \\
18   &    $(0,0,\delta,0, 0,\delta)$  & $\left(\frac{C_{33}}{2}+C_{36}+\frac{C_{66}}{2}\right) \delta^{2}$ \\
19   &    $(0,0,0,\delta, \delta,0)$  & $\left(\frac{C_{44}}{2}+C_{45}+\frac{C_{55}}{2}\right) \delta^{2}$ \\
20   &    $(0,0,0,\delta, 0,\delta)$  & $\left(\frac{C_{44}}{2}+C_{46}+\frac{C_{66}}{2}\right) \delta^{2}$ \\
21   &    $(0,0,0,0,\delta, \delta)$  & $\left(\frac{C_{55}}{2}+C_{56}+\frac{C_{66}}{2}\right) \delta^{2}$ \\

\hline
\end{tabular}
\end{table*}

%%%%%%%%%%%%%%%%%%%%%%%%%%%%%%%%%%%%%%%%%%%%%%%%%%%%%%%%%%%%%%%%%%%%%%%%%%%%%%%%%%%%%%%%%%%%%%%%%%%%
\begin{description}
\item [{2. Monoclinic System (Space group numbers: 3-15)}]~
\end{description}

There are 13 independent elastic constants: $C_{11}$, $C_{12}$, $C_{13}$, $C_{15}$, $C_{22}$, $C_{23}$,  $C_{25}$, $C_{33}$, $C_{35}$, $C_{44}$, $C_{46}$, $C_{55}$ and $C_{66}$

The elastic stiffness tensor matrix is expressed by
\begin{equation}
C_{ij}=\left(\begin{array}{cccccc}
C_{11} & C_{12} & C_{13} & 0 &  C_{15} & 0 \\
C_{12} & C_{22} & C_{23} & 0  & C_{25} & 0  \\
C_{13} & C_{23} & C_{33} & 0  & C_{35} & 0\\
0 & 0 & 0 & C_{44} & 0 & C_{46} \\
C_{15} & C_{25} & C_{35} & 0 & C_{55} & 0 \\
0 & 0 & 0 & 0 & C_{46} & C_{66}
\end{array}\right).
\end{equation}

\begin{table*}[htbp]
\centering
\caption{List of strain modes and the derived elastic constants for \textbf{monoclinic} system used in VASPKIT based on energy-strain approach.}
\begin{tabular}{ccc}
\hline
Strain index & Strain vector $\boldsymbol{\varepsilon}$ & Elastic energy $\frac{\Delta E}{V}$  \\
\hline
1   &    $(\delta, 0, 0, 0, 0, 0)$       &      $\frac{1}{2} C_{11} \delta^{2}$ \\
2   &    $(0, \delta, 0, 0, 0, 0)$       &      $\frac{1}{2} C_{22} \delta^{2}$ \\
3   &    $(0, 0, \delta, 0, 0, 0)$       &      $\frac{1}{2} C_{33} \delta^{2}$ \\
4   &    $( 0, 0, 0,\delta, 0, 0)$       &      $\frac{1}{2} C_{44} \delta^{2}$ \\
5   &    $(0, 0, 0, 0,\delta,  0)$       &      $\frac{1}{2} C_{55} \delta^{2}$ \\
6   &    $(0, 0, 0, 0, 0,\delta )$       &      $\frac{1}{2} C_{66} \delta^{2}$ \\
7   &    $(\delta, \delta,0,0,0,0)$  & $\left(\frac{C_{11}}{2}+C_{12}+\frac{C_{22}}{2}\right) \delta^{2}$ \\
8   &    $(\delta,0, \delta,0,0,0)$  & $\left(\frac{C_{11}}{2}+C_{13}+\frac{C_{33}}{2}\right) \delta^{2}$ \\
9   &    $(\delta,0,0,0,\delta,0)$  & $\left(\frac{C_{11}}{2}+C_{15}+\frac{C_{55}}{2}\right) \delta^{2}$ \\
10   &    $(0,\delta, \delta,0,0,0)$  & $\left(\frac{C_{22}}{2}+C_{23}+\frac{C_{33}}{2}\right) \delta^{2}$ \\
11   &    $(0,\delta,0,0,\delta,0)$  & $\left(\frac{C_{22}}{2}+C_{25}+\frac{C_{55}}{2}\right) \delta^{2}$ \\
12   &    $(0,0,\delta,0,\delta,0)$  & $\left(\frac{C_{33}}{2}+C_{35}+\frac{C_{55}}{2}\right) \delta^{2}$ \\
13   &    $(0,0,0,\delta, 0,\delta)$  & $\left(\frac{C_{44}}{2}+C_{46}+\frac{C_{66}}{2}\right) \delta^{2}$ \\
\hline
\end{tabular}
\end{table*}

%%%%%%%%%%%%%%%%%%%%%%%%%%%%%%%%%%%%%%%%%%%%%%%%%%%%%%%%%%%%%%%%%%%%%%%%%%%%%%%%%%%%%%%%%%%%%%%%%%%%
\begin{description}
\item [{3. Orthorhombic System (Space group numbers: 16-74)}]~
\end{description}

There are 9 independent elastic constants: $C_{11}$, $C_{12}$, $C_{13}$, $C_{22}$, $C_{23}$, $C_{33}$, $C_{44}$, $C_{55}$ and $C_{66}$  
 
The elastic stiffness tensor matrix is expressed by
\begin{equation}
C_{ij}=\left(\begin{array}{cccccc}
C_{11} & C_{12} & C_{13} & 0 & 0 & 0 \\
C_{12} & C_{22} & C_{23} & 0 & 0 & 0 \\
C_{13} & C_{23} & C_{33} & 0 & 0 & 0 \\
0 & 0 & 0 & C_{44} & 0 & 0 \\
0 & 0 & 0 & 0 & C_{55} & 0 \\
0 & 0 & 0 & 0 & 0 & C_{66}
\end{array}\right).
\end{equation}

\begin{table*}[htbp]
\centering
\caption{List of strain modes and the derived elastic constants for \textbf{orthorhombic} system used in VASPKIT based on energy-strain approach.}
\begin{tabular}{ccc}
\hline
Strain index & Strain vector $\boldsymbol{\varepsilon}$ & Elastic energy $\frac{\Delta E}{V}$  \\
\hline
1   &    $(\delta, 0, 0, 0, 0, 0)$       &      $\frac{1}{2} C_{11} \delta^{2}$ \\
2   &    $(0, \delta, 0, 0, 0, 0)$       &      $\frac{1}{2} C_{22} \delta^{2}$ \\
3   &    $(0, 0, \delta, 0, 0, 0)$       &      $\frac{1}{2} C_{33} \delta^{2}$ \\
4   &    $( 0, 0, 0,\delta, 0, 0)$       &      $\frac{1}{2} C_{44} \delta^{2}$ \\
5   &    $(0, 0, 0, 0,\delta,  0)$       &      $\frac{1}{2} C_{55} \delta^{2}$ \\
6   &    $(0, 0, 0, 0, 0,\delta )$       &      $\frac{1}{2} C_{66} \delta^{2}$ \\
7   &    $(\delta, \delta,0,0,0,0)$  & $\left(\frac{C_{11}}{2}+C_{12}+\frac{C_{22}}{2}\right) \delta^{2}$ \\
8   &    $(\delta,0, \delta,0,0,0)$  & $\left(\frac{C_{11}}{2}+C_{13}+\frac{C_{33}}{2}\right) \delta^{2}$ \\
9   &    $(0,\delta, \delta,0,0,0)$  & $\left(\frac{C_{22}}{2}+C_{23}+\frac{C_{33}}{2}\right) \delta^{2}$ \\
\hline
\end{tabular}
\end{table*}

%%%%%%%%%%%%%%%%%%%%%%%%%%%%%%%%%%%%%%%%%%%%%%%%%%%%%%%%%%%%%%%%%%%%%%%%%%%%%%%%%%%%%%%%%%%%%%%%%%%%
\begin{description}
\item [{4. Tetragonal II System (Space group numbers: 75-88)}]~
\end{description}

There are 7 independent elastic constants: $C_{11}$, $C_{12}$, $C_{13}$, $C_{16}$, $C_{33}$, $C_{44}$ and $C_{66}$  

The elastic stiffness tensor matrix is expressed by
\begin{equation}
C_{ij}=\left(\begin{array}{cccccc}
C_{11} & C_{12} & C_{13} & 0 & 0 & C_{16} \\
C_{12} & C_{11} & C_{13} & 0 & 0 & -C_{16} \\
C_{13} & C_{13} & C_{33} & 0 & 0 & 0 \\
0 & 0 & 0 & C_{44} & 0 & 0 \\
0 & 0 & 0 & 0 & C_{44} & 0 \\
C_{16} & -C_{16} & 0 & 0 & 0 & C_{66}
\end{array}\right).
\end{equation}

\begin{table*}[htbp]
\centering
\caption{List of strain modes and the derived elastic constants for \textbf{tetragonal II} system used in VASPKIT based on energy-strain approach.}
\begin{tabular}{ccc}
\hline
Strain index & Strain vector $\boldsymbol{\varepsilon}$ & Elastic energy $\frac{\Delta E}{V}$  \\
\hline
1   &    $(\delta, \delta, 0, 0, 0, 0)$       &      $(C_{11}+C_{12}) \delta^{2}$ \\
2   &    $(0,0,0,0,0,\delta)$     & $\frac{1}{2} C_{66} \delta^{2}$ \\
3   &    $(0,0,\delta, 0, 0, 0 )$  & $\frac{1}{2} C_{33} \delta^{2}$ \\
4   &    $(0,0,0,\delta,\delta, 0 )$  & $C_{44} \delta^{2}$ \\
5   &    $(\delta, \delta, \delta, 0, 0, 0 )$  & $\left(C_{11}+C_{12}+2 C_{13}+\frac{C_{33}}{2}\right) \delta^{2}$ \\
6   &    $(0,\delta, \delta,0,0,0)$  & $\left(\frac{C_{11}}{2}+C_{13}+\frac{C_{33}}{2}\right) \delta^{2}$ \\
7   &    $(\delta, 0,0,0,0,\delta)$  & $\left(\frac{C_{11}}{2}+C_{16}+\frac{C_{66}}{2}\right) \delta^{2}$ \\
\hline
\end{tabular}
\end{table*}

%%%%%%%%%%%%%%%%%%%%%%%%%%%%%%%%%%%%%%%%%%%%%%%%%%%%%%%%%%%%%%%%%%%%%%%%%%%%%%%%%%%%%%%%%%%%%%%%%%%%
\begin{description}
\item [{5. Tetragonal I System (Space group numbers: 89-142)}]~
\end{description}

There are 6 independent elastic constants: $C_{11}$, $C_{12}$, $C_{13}$, $C_{33}$, $C_{44}$ and $C_{66}$  

The elastic stiffness tensor matrix is expressed by
\begin{equation}
C_{ij}=\left(\begin{array}{cccccc}
C_{11} & C_{12} & C_{13} & 0 & 0 & 0 \\
C_{12} & C_{11} & C_{13} & 0 & 0 & 0 \\
C_{13} & C_{13} & C_{33} & 0 & 0 & 0 \\
0 & 0 & 0 & C_{44} & 0 & 0 \\
0 & 0 & 0 & 0 & C_{44} & 0 \\
0 & 0 & 0 & 0 & 0 & C_{66}
\end{array}\right).
\end{equation}

\begin{table*}[htbp]
\centering
\caption{List of strain modes and the derived elastic constants for \textbf{tetragonal I} system used in VASPKIT based on energy-strain approach.}
\begin{tabular}{ccc}
\hline
Strain index & Strain vector $\boldsymbol{\varepsilon}$ & Elastic energy $\frac{\Delta E}{V}$  \\
\hline
1   &    $(\delta, \delta, 0, 0, 0, 0)$       &      $(C_{11}+C_{12}) \delta^{2}$ \\
2   &    $(0,0,0,0,0,\delta)$     & $\frac{1}{2} C_{66} \delta^{2}$ \\
3   &    $(0,0,\delta, 0, 0, 0 )$  & $\frac{1}{2} C_{33} \delta^{2}$ \\
4   &    $(0,0,0,\delta,\delta, 0 )$  & $C_{44} \delta^{2}$ \\
5   &    $(\delta, \delta, \delta, 0, 0, 0 )$  & $\left(C_{11}+C_{12}+2 C_{13}+\frac{C_{33}}{2}\right) \delta^{2}$ \\
6   &    $(0,\delta, \delta,0,0,0)$  & $\left(\frac{C_{11}}{2}+C_{13}+\frac{C_{33}}{2}\right) \delta^{2}$ \\
\hline
\end{tabular}
\end{table*}

%%%%%%%%%%%%%%%%%%%%%%%%%%%%%%%%%%%%%%%%%%%%%%%%%%%%%%%%%%%%%%%%%%%%%%%%%%%%%%%%%%%%%%%%%%%%%%%%%%%%
\begin{description}
\item [{6. Trigonal II System (Space group numbers: 143-148)}]~
\end{description}

There are 7 independent elastic constants: $C_{11}$, $C_{12}$, $C_{13}$, $C_{14}$, $C_{15}$, $C_{33}$ and $C_{44}$ 

The elastic stiffness tensor matrix is expressed by
\begin{equation}
C_{ij}=\left(\begin{array}{cccccc}
C_{11} & C_{12} & C_{13} & C_{14} & C_{15} & 0 \\
C_{12} & C_{11} & C_{13} & -C_{14} & -C_{15} & 0 \\
C_{13} & C_{13} & C_{33} & 0 & 0 & 0 \\
C_{14} & -C_{14} & 0 & C_{44} & 0 & -C_{15} \\
C_{15} & -C_{15} & 0 & 0 & C_{44} & C_{14} \\
0 & 0 & 0 & -C_{15} & C_{14} & \frac{C_{11}-C_{12}}{2}
\end{array}\right).
\end{equation}

\begin{table*}[htbp]
\centering
\caption{List of strain modes and the derived elastic constants for \textbf{trigonal II} system used in VASPKIT based on energy-strain approach.}
\begin{tabular}{ccc}
\hline
Strain index & Strain vector $\boldsymbol{\varepsilon}$ & Elastic energy $\frac{\Delta E}{V}$  \\
\hline
1   &    $(\delta, \delta, 0, 0, 0, 0)$       &      $(C_{11}+C_{12}) \delta^{2}$ \\
2   &    $(0,0,0,0,0,\delta)$     & $\frac{1}{4}\left(C_{11}-C_{12}\right) \delta^{2}$ \\
3   &    $(0,0,\delta, 0, 0, 0 )$  & $\frac{1}{2} C_{33} \delta^{2}$ \\
4   &    $(0,0,0,\delta,\delta, 0 )$  & $C_{44} \delta^{2}$ \\
5   &    $(\delta, \delta, \delta, 0, 0, 0 )$  & $\left(C_{11}+C_{12}+2 C_{13}+\frac{C_{33}}{2}\right) \delta^{2}$ \\
6   &    $(0,0,0,0,\delta, \delta)$  & $\left(\frac{C_{11}}{4}-\frac{C_{12}}{4}+C_{14}+\frac{C_{44}}{2}\right) \delta^{2}$ \\
7   &    $(0,0,0,\delta,0, \delta)$  & $\left(\frac{C_{11}}{4}-\frac{C_{12}}{4}-C_{15}+\frac{C_{44}}{2}\right) \delta^{2}$ \\
\hline
\end{tabular}
\end{table*}

%%%%%%%%%%%%%%%%%%%%%%%%%%%%%%%%%%%%%%%%%%%%%%%%%%%%%%%%%%%%%%%%%%%%%%%%%%%%%%%%%%%%%%%%%%%%%%%%%%%%
\begin{description}
\item [{7. Trigonal I System (Space group numbers: 149-167)}]~
\end{description}

There are 6 independent elastic constants: $C_{11}$, $C_{12}$, $C_{13}$, $C_{14}$, $C_{33}$ and $C_{44}$ 

The elastic stiffness tensor matrix is expressed by
\begin{equation}
C_{ij}=\left(\begin{array}{cccccc}
C_{11} & C_{12} & C_{13} & C_{14} & 0 & 0 \\
C_{12} & C_{11} & C_{13} & -C_{14} & 0 & 0 \\
C_{13} & C_{13} & C_{33} & 0 & 0 & 0 \\
C_{14} & -C_{14} & 0 & C_{44} & 0 & 0 \\
0 & 0 & 0 & 0 & C_{44} & C_{14} \\
0 & 0 & 0 & 0 & C_{14} & \frac{C_{11}-C_{12}}{2}
\end{array}\right).
\end{equation}

\begin{table*}[htbp]
\centering
\caption{List of strain modes and the derived elastic constants for \textbf{trigonal I} system used in VASPKIT based on energy-strain approach.}
\begin{tabular}{ccc}
\hline
Strain index & Strain vector $\boldsymbol{\varepsilon}$ & Elastic energy $\frac{\Delta E}{V}$  \\
\hline
1   &    $(\delta, \delta, 0, 0, 0, 0)$       &      $(C_{11}+C_{12}) \delta^{2}$ \\
2   &    $(0,0,0,0,0,\delta)$     & $\frac{1}{4}\left(C_{11}-C_{12}\right) \delta^{2}$ \\
3   &    $(0,0,\delta, 0, 0, 0 )$  & $\frac{1}{2} C_{33} \delta^{2}$ \\
4   &    $(0,0,0,\delta,\delta, 0 )$  & $C_{44} \delta^{2}$ \\
5   &    $(\delta, \delta, \delta, 0, 0, 0 )$  & $\left(C_{11}+C_{12}+2 C_{13}+\frac{C_{33}}{2}\right) \delta^{2}$ \\
6   &    $(0,0,0,0,\delta, \delta)$  & $\left(\frac{C_{11}}{4}-\frac{C_{12}}{4}+C_{14}+\frac{C_{44}}{2}\right) \delta^{2}$ \\
\hline
\end{tabular}
\end{table*}

%%%%%%%%%%%%%%%%%%%%%%%%%%%%%%%%%%%%%%%%%%%%%%%%%%%%%%%%%%%%%%%%%%%%%%%%%%%%%%%%%%%%%%%%%%%%%%%%%%%% 
\begin{description}
\item [{8. Hexagonal System (Space group numbers: 168–194)}]~
\end{description}

There are 5 independent elastic constants: $C_{11}$, $C_{12}$, $C_{13}$, $C_{33}$ and $C_{44}$

The elastic stiffness tensor matrix is expressed by
\begin{equation}
C_{ij}=\left(\begin{array}{cccccc}
C_{11} & C_{12} & C_{13} & 0 & 0 & 0 \\
C_{12} & C_{11} & C_{13} & 0 & 0 & 0 \\
C_{13} & C_{13} & C_{33} & 0 & 0 & 0 \\
0 & 0 & 0 & C_{44} & 0 & 0 \\
0 & 0 & 0 & 0 & C_{44} & 0 \\
0 & 0 & 0 & 0 & 0 & \frac{C_{11}-C_{12}}{2}
\end{array}\right).
\end{equation}

\begin{table*}[htbp]
\centering
\caption{List of strain modes and the derived elastic constants for \textbf{hexagonal} system used in VASPKIT based on energy-strain approach.}
\begin{tabular}{ccc}
\hline
Strain index & Strain vector $\boldsymbol{\varepsilon}$ & Elastic energy $\frac{\Delta E}{V}$  \\
\hline
1   &    $(\delta, \delta, 0, 0, 0, 0)$       &      $(C_{11}+C_{12}) \delta^{2}$ \\
2   &    $(0,0,0,0,0,\delta)$     & $\frac{1}{4}\left(C_{11}-C_{12}\right) \delta^{2}$ \\
3   &    $(0,0,\delta, 0, 0, 0 )$  & $\frac{1}{2} C_{33} \delta^{2}$ \\
4   &    $(0,0,0,\delta,\delta, 0 )$  & $C_{44} \delta^{2}$ \\
5   &    $(\delta, \delta, \delta, 0, 0, 0 )$  & $\left(C_{11}+C_{12}+2 C_{13}+\frac{C_{33}}{2}\right) \delta^{2}$ \\
\hline
\end{tabular}
\end{table*}

%%%%%%%%%%%%%%%%%%%%%%%%%%%%%%%%%%%%%%%%%%%%%%%%%%%%%%%%%%%%%%%%%%%%%%%%%%%%%%%%%%%%%%%%%%%%%%%%%%%%
\begin{description}
\item [{9. Cubic System (Space group numbers: 195–230)}]~
\end{description}

There are 3 independent elastic constants: $C_{11}$, $C_{12}$ and $C_{44}$

The elastic stiffness tensor matrix is expressed by

\begin{equation}
\label{Cij_cubic}
C_{ij}=\left(\begin{array}{cccccc}
C_{11} & C_{12} & C_{12} & 0 & 0 & 0 \\
C_{12} & C_{11} & C_{12} & 0 & 0 & 0 \\
C_{12} & C_{12} & C_{11} & 0 & 0 & 0 \\
0 & 0 & 0 & C_{44} & 0 & 0 \\
0 & 0 & 0 & 0 & C_{44} & 0 \\
0 & 0 & 0 & 0 & 0 & C_{44}
\end{array}\right).
\end{equation}

\begin{table*}[htbp]
\centering
\caption{List of strain modes and the derived elastic constants for \textbf{cubic} system used in VASPKIT based on energy-strain approach.}
\begin{tabular}{ccc}
\hline
Strain index & Strain vector $\boldsymbol{\varepsilon}$ & Elastic energy $\frac{\Delta E}{V}$  \\
\hline
1   &    $(0,0,0, \delta, \delta, \delta)$     & $\frac{3}{2} C_{44} \delta^{2}$ \\
2   &    $(\delta, \delta, 0, 0, 0, 0)$       &      $(C_{11}+C_{12}) \delta^{2}$ \\
3   &    $(\delta, \delta, \delta, 0, 0, 0 )$  & $\frac{3}{2}(C_{11}+2C_{12}) \delta^{2}$ \\
\hline
\end{tabular}
\end{table*}

\section{Elastic stiffness tensor matrix and strain modes  for 2D crystal systems}
%%%%%%%%%%%%%%%%%%%%%%%%%%%%%%%%%%%%%%%%%%%%%%%%%%%%%%%%%%%%%%%%%%%%%%%%%%%%%%%%%%%%%%%%%%%%%%%%%%%%
\begin{description}
\item [{1. 2D Oblique System}]~
\end{description}

There are 6 independent elastic constants: $C_{11}$, $C_{12}$, $C_{16}$, $C_{22}$, $C_{26}$ and $C_{66}$
\begin{equation}
C_{ij}=\left(\begin{array}{ccc}
C_{11} & C_{12} & C_{16} \\
C_{21} & C_{22} & C_{26} \\
C_{61} & C_{62} & C_{66}
\end{array}\right)
\end{equation}

\begin{table*}[htbp]
\centering
\caption{List of strain modes and the derived elastic constants for 2D \textbf{oblique} system used in VASPKIT based on energy-strain approach.}
\begin{tabular}{ccc}
\hline
Strain index & Strain vector $\boldsymbol{\varepsilon}$ & Elastic energy $\frac{\Delta E}{V}$  \\
\hline
1   &    $(\delta, 0, 0, 0, 0, 0)$       &      $\frac{1}{2} C_{11} \delta^{2}$ \\
2   &    $(0, \delta, 0, 0, 0, 0)$       &      $\frac{1}{2} C_{22} \delta^{2}$ \\
3   &    $(0, 0, 0, 0, 0,\delta )$       &      $\frac{1}{2} C_{66} \delta^{2}$ \\
4   &    $(\delta, \delta,0,0,0,0)$  & $\left(\frac{C_{11}}{2}+C_{12}+\frac{C_{22}}{2}\right) \delta^{2}$ \\
5   &    $(\delta,0,0,0,0,\delta)$  & $\left(\frac{C_{11}}{2}+C_{16}+\frac{C_{66}}{2}\right) \delta^{2}$ \\
6   &    $(0,\delta,0,0,0,\delta)$  & $\left(\frac{C_{22}}{2}+C_{26}+\frac{C_{66}}{2}\right) \delta^{2}$ \\
\hline
\end{tabular}
\end{table*}

%%%%%%%%%%%%%%%%%%%%%%%%%%%%%%%%%%%%%%%%%%%%%%%%%%%%%%%%%%%%%%%%%%%%%%%%%%%%%%%%%%%%%%%%%%%%%%%%%%%%
\begin{description}
\item [{2. 2D Rectangular System}]~
\end{description}

There are 4 independent elastic constants: $C_{11}$, $C_{12}$, $C_{22}$ and $C_{66}$
\begin{equation}
C_{ij}=\left(\begin{array}{ccc}
C_{11} & C_{12} &  0 \\
C_{21} & C_{22} & 0 \\
0   & 0 & C_{66}  
\end{array}\right)
\end{equation}

\begin{table*}[htbp]
\centering
\caption{List of strain modes and the derived elastic constants for 2D \textbf{rectangular} system used in VASPKIT based on energy-strain approach.}
\begin{tabular}{ccc}
\hline
Strain index & Strain vector $\boldsymbol{\varepsilon}$ & Elastic energy $\frac{\Delta E}{V}$  \\
\hline
1   &    $(\delta, 0, 0, 0, 0, 0)$       &      $\frac{1}{2} C_{11} \delta^{2}$ \\
2   &    $(0, \delta, 0, 0, 0, 0)$       &      $\frac{1}{2} C_{22} \delta^{2}$ \\
3   &    $(0, 0, 0, 0, 0,\delta )$       &      $\frac{1}{2} C_{66} \delta^{2}$ \\
4   &    $(\delta, \delta,0,0,0,0)$  & $\left(\frac{C_{11}}{2}+C_{12}+\frac{C_{22}}{2}\right) \delta^{2}$ \\
\hline
\end{tabular}
\end{table*}

%%%%%%%%%%%%%%%%%%%%%%%%%%%%%%%%%%%%%%%%%%%%%%%%%%%%%%%%%%%%%%%%%%%%%%%%%%%%%%%%%%%%%%%%%%%%%%%%%%%%
\begin{description}
\item [{3. 2D Square System}]~
\end{description}

There are 3 independent elastic constants: $C_{11}$, $C_{12}$ and $C_{66}$
\begin{equation}
C_{ij}=\left(\begin{array}{ccc}
C_{11} & C_{12} &  0 \\
C_{21} & C_{11} & 0 \\
0   & 0 & C_{66}  
\end{array}\right)
\end{equation}

\begin{table*}[htbp]
\centering
\caption{List of strain modes and the derived elastic constants for 2D \textbf{square} system used in VASPKIT based on energy-strain approach.}
\begin{tabular}{ccc}
\hline
Strain index & Strain vector $\boldsymbol{\varepsilon}$ & Elastic energy $\frac{\Delta E}{V}$  \\
\hline
1   &    $(\delta, 0, 0, 0, 0, 0)$       &      $\frac{1}{2} C_{11} \delta^{2}$ \\
2   &    $(0, 0, 0, 0, 0,\delta )$       &      $\frac{1}{2} C_{66} \delta^{2}$ \\
3   &    $(\delta, \delta,0,0,0,0)$  & $\left(C_{11}+C_{12}\right) \delta^{2}$ \\
\hline
\end{tabular}
\end{table*}

%%%%%%%%%%%%%%%%%%%%%%%%%%%%%%%%%%%%%%%%%%%%%%%%%%%%%%%%%%%%%%%%%%%%%%%%%%%%%%%%%%%%%%%%%%%%%%%%%%%%
\begin{description}
\item [{4. 2D Hexagonal System}]~
\end{description}

There are 2 independent elastic constants: $C_{11}$ and $C_{12}$
\begin{equation}
C_{ij}=\left(\begin{array}{ccc}
C_{11} & C_{12} &  0 \\
C_{21} & C_{11} & 0 \\
0   & 0 & \frac{C_{11}-C_{12}}{2}  
\end{array}\right)
\end{equation}

\begin{table*}[htbp]
\centering
\caption{List of strain modes and the derived elastic constants for 2D \textbf{hexagonal} system used in VASPKIT based on energy-strain approach.}
%\caption{strain modes and the corresponding strain vector expressed in the Voigt notation, that are used by VASPKIT in the energy-strain approach.}
\begin{tabular}{ccc}
\hline
Strain index & Strain vector $\boldsymbol{\varepsilon}$ & Elastic energy $\frac{\Delta E}{V}$  \\
\hline
1   &    $(\delta, 0, 0, 0, 0, 0)$       &      $\frac{1}{2} C_{11} \delta^{2}$ \\
2   &    $(\delta, \delta,0,0,0,0)$  & $\left(C_{11}+C_{12}\right) \delta^{2}$ \\
\hline
\end{tabular}
\end{table*}

\bibliographystyle{elsarticle-num}
%\bibliography{vaspkit}

\end{document}